# Change in Magnetic Order in NiPS$_3$ Single Crystals Induced by a Molecular Intercalation

Nirman Chakraborty,[1] Adi Harchol,[1] Azhar Abu-Hariri,[1] Rajesh Kumar Yadav,[2,3] Muhamed Dawod,[4] Diksha Prabhu Gaonkar,[1] Kusha Sharma,[1] Anna Eyal,[5] Yaron Amouyal,[4] Doron Naveh[2,3] and Efrat Lifshitz[1*]

[1]*Schulich Faculty of Chemistry, Solid State Institute, Russel Berrie Nanotechnology Institute, Grand Program for Energy and the Helen Diller Quantum Center, Technion-Israel Institute of Technology, Haifa 3200003, Israel*

[2]*Faculty of Engineering, Bar-Ilan University, Ramat-Gan 5290002, Israel*

[3]*Institute of Nanotechnology and Advanced Materials, Bar-Ilan University, Ramat-Gan 5290002, Israel*

[4]*Department of Materials Science and Engineering, Technion-Israel Institute of Technology, Haifa 3200003, Israel*

[5]*Department of Physics, Technion, Haifa 3200003, Israel*

*Corresponding author: ssefrat@technion.ac.il

**Abstract**

Intercalation is a robust method for tuning the physical properties of a vast number of van der Waals (vdW) materials. However, the prospects of using intercalation to modify magnetism in vdWs systems and the associated mechanisms have not been investigated adequately. In this work, we modulate magnetic order in an XY antiferromagnet NiPS$_3$ single crystals by introducing pyridine molecules into the vdW's gap under different thermal conditions. X-ray diffraction measurements indicated pronounced changes in the lattice parameter beta (*β*), while magnetization measurements at in-plane and out-of-plane configurations exposed reversal trends in the crystals' Néel temperatures through intercalation/de-intercalation processes. The changes in magnetic ordering were also supported by three-dimensional thermal diffusivity experiments. The preferred orientation of the pyridine dipoles within vdW gaps was deciphered *via* polarized Raman spectroscopy. The results highlight the relation between the preferential alignment of the intercalants, thermal transport, and crystallographic disorder along with the modulation of anisotropy in the magnetic order. The theoretical concept of double-exchange interaction in NiPS$_3$ was employed to explain the intercalation-induced magnetic ordering. The study uncovers the merit of intercalation as a foundation for spin switches and spin transistors in advanced quantum devices.

**Introduction**

Intercalation is a unique strategy to modulate the physicochemical properties of van der Waals' materials, which can improve their utility in rechargeable batteries, catalysis, optoelectronics, and magnetic devices [1,2]. The interlayer gap in lamellar materials becomes a perfect platform to accommodate host-guest interactions, which can introduce interlayer exchange interactions in otherwise weakly interacting atomic layers [3,4]. The introduction of foreign electronic entities in a stable electronic structure can be perceived as a perturbation that eventually leads to intrinsic electronic and magnetic adjustments, resulting in the emergence of new and unique physical properties. Since the discovery of magnetism in vdW systems, studies have been conducted to investigate the effects of intercalation in both antiferromagnetic (AFM) and ferromagnetic (FM) vdW materials [5-7]. While these studies in FM vdW systems are relatively new, intercalation has been tried in a varied class of AFM vdW materials [7]. Among the AFM vdWs, members of the transition metal phosphorus trichalcogenide $MPX_3$ (M- first-row transition metals, P-phosphorous, X-S, Se) family have been used as hosts for intercalating different molecular and ionic species [8-12]. A single layer in the $MPX_3$ system consists of transition metal ions M (usually in the coordination state 2), octahedrally surrounded by chalcogenide atoms (S, Se), which are interconnected via $(P_2X_6)^{4-}$ units [13], thus the chemical formula often written as $M_2P_2X_6$. The metal ions with high spin configuration across the layers are arranged in a honeycomb pattern with P2 atoms located in the hexagon centers. This hexagonal framework enables the AFM spin alignment of the individual layers, either in perpendicular (Ising or isotropic Heisenberg models) or parallel (anisotropic Heisenberg model or XY model) orientation, with respect to the basal plane [14-16].

Depending on the choice of the transition metal and X atom (S or Se), this class of materials has been found to exhibit diverse anti-ferromagnetism (AFM) [1]. With the intralayer arrangement controlled by strong spin-spin exchange interactions among M-M atoms, the weak spin interactions across the vdW gap endow an inherent anisotropy [8,10]. Owing to novel properties like low stray magnetic fields and THz spin-flip frequency, these two-dimensional (2D) AFM materials stand as potential candidates for spin filters, spin transistors, and photodetectors [17-19]. $NiPS_3$ is a semiconducting XY antiferromagnet with AFM-paramagnetic phase transition at a Néel temperature of 150 K [8,13]. Due to the high tendency towards intercalation, intercalates of different types have been introduced into the $NiPS_3$ systems for specific purposes [8, 20-22]. Various organic ion intercalates have been found to induce AFM to ferrimagnetic (FIM) or FM transitions in NiPS$_3$ by electron transfer to the host [20]. Electrochemical intercalation of tetrabutylammonium cations results in a FIM hybrid compound displaying a transition temperature of 78 K [20]. Intercalation of 1,10-phenanthroline into NiPS$_3$ *via* iron dopant seeding has been found to suppress AFM and gave rise to FIM at a temperature around 75 K [21]. Electrochemical intercalation of Li into NiPS$_3$ has been found to reduce its magnetic susceptibility and led to the emergence of FIM at low temperatures [22]. Electrochemical intercalation of ionic compounds like 1-Ethyl-3-methylimidazolium chloride (EMIM)-BF$_4$ in NiPS$_3$ has been found to bring about nearly isotropic spin fluctuations [8]. While most reported studies have been able to achieve the transformation of pristine NiPS$_3$ from one magnetic state to another *via* electron transfer from intercalate, there have been no significant attempts to explain the structural and electronic property modifications in these systems induced by the intercalate type and the intercalation processes. This brings into the picture the necessity of investigating the above aspects of the layered systems that help accommodate a specific intercalation strategy and, consequently, a typical magnetic ordering.

NiPS$_3$ exhibits an AFM-zigzag ordering in its magnetic ground state [23]. Each Ni$^{2+}$ (3$d^8$, S=1) spin is coupled ferromagnetically to two of the nearest neighbors (NNs) and antiferromagnetically to the third, such that within the layer, the Ni$^{2+}$ moments appear as ferromagnetic chains coupled antiferromagnetically to each other [23]. In interpreting the magnetic properties, two factors play a vital role: (i) the 2D nature of magnetic interactions (both direct and super-exchange) [24] and (ii) the trigonal distortion of the MS$_6$ octahedra due to the interaction of the orbital state of M and the surrounding crystalline field (single ion anisotropy) (see Figure 1(a)) [13]. Neutron diffraction studies on NiPS$_3$ have revealed the orientation of the spins in the "*ab*" crystallographic plane, the magnetization axis being perpendicular to the trigonal axis of the unit cell [25-27]. The Heisenberg Hamiltonian for magnetic interaction between neighboring atoms in the absence of lattice distortions and other perturbations can be written as:

$$H = -2\sum\{J_{\parallel}(S_{ix}S_{jx}+S_{iy}S_{jy}) + J_{\perp}(S_{iz}S_{jz})\}\ldots\ldots\ldots\ldots(1)$$

where the summation is over all pairs of spins in the lattice (all permutations and combinations of nearest neighbor interaction) [13,28]. $J_{\perp}$ and $J_{\parallel}$ represent the perpendicular and parallel (concerning the "*ab*" plane) exchange interactions between the spins $S_i$ and $S_j$. In the ideal case, for NiPS$_3$, $J_\perp$=0. Considering the effect of axial distortion, the term contributing to single-ion anisotropy can be added to the Hamiltonian as:

$$H = -2\sum\{J_\parallel (S_{ix}S_{jx}+S_{iy}S_{jy}) + J_\perp (S_{iz}S_{jz})\} + DS_{iz}^2 \ldots\ldots\ldots\ldots\ldots\ldots\ldots\ldots\ldots(2)$$

where the quadratic axial crystal-field parameter $D$ arises from the combined effects of the crystal-field and spin-orbit coupling (SOC).

Modulation of magnetic anisotropy in XY systems requires a gradual strengthening of the $J_\perp$ component in equation 2 at the expense of $J_\parallel$. This can be done by modifying the nature of in-plane spin-exchange interactions to generate prominent out-of-plane spin components. Intercalation of FM vdWs by ionic species has been reported to induce electron transfer from the intercalate to the host material, enabling direct-exchange interactions between M-M doublet $d$ orbitals [29-31]. The mode of accommodation of incoming electrons determines the strength of direct exchange between the M-M atoms, which can alter the FM-paramagnetic phase transition behavior and flipping direction of in-plane spin components [29-31]. This, in addition to the increment in Hund's coupling strength, leads to the occurrence of M-M double-exchange interactions, delineating the plausible physical origin of modulation of magnetic order in 2D systems *via* intercalation [29]. However, small ionic species can tend to accommodate themselves in the octahedral sites of the vdW gaps [22], making the conventional thermal routes inefficient for total deintercalation. This requires the introduction of a heterocyclic compound with a size comparable to the vdW gap, with the capability to efficiently transfer electrons to the host [32].

Based on the above concept, our work introduces a reversible tailoring of the magnetic anisotropy in the XY antiferromagnet NiPS$_3$ by incorporating pyridine molecules within the vdWs gaps under different thermal conditions (see Figure 1) [32]. Single crystal X-ray diffraction probing reveals the long-range structural details of the pristine and the intercalated NiPS$_3$ systems, including identifying the defects and disorders post-intercalation [33]. Microstructural studies into the role of intercalates in generating structural disorders have been performed using high-angle annular dark-field scanning transmission electron microscopic experiments on cross-sections of pristine and intercalated single crystals. X-ray Photoelectron Spectroscopy and *in-situ* polarized Raman spectroscopy have been employed to investigate the preferential orientation of intercalates within the vdW gaps [34,35]. Magnetic and thermal property measurements followed the change in Néel temperature, magnetic anisotropies, and thermal properties upon intercalation. An understanding of the change in magnetic and thermal properties in terms of the short-range and long-range structural modifications, accompanied by the phenomenon of double-exchange interactions, unveiled the interplay of intrinsic and extrinsic disorders in tuning the interfacial spin properties. The above methodologies uncovered the mechanisms of intercalation in $NiPS_3$ and similarly can be implemented for other members of the $MPX_3$ family as well.

**Results and discussion**

**Materials processing and composition/structural characterizations.** Bulk single crystals of $NiPS_3$ were formed using a chemical vapor transport (CVT) technique [8]. The pyridine intercalation was achieved *via* a soaking procedure in a chemical solution under variable temperatures and inert conditions [32]. The examined samples are labeled hereon according to the intercalation temperature as PY_X (X = 25, 40, 50, 65, 75°C). A deintercalation process was performed by a thermal treatment in vacuum. The crystal structures of the pristine and intercalated crystals were verified by Single crystal X-ray diffraction (SXRD) measurements. The compositions were revealed from scanning electron microscopy combined with energy dispersive X-ray (SEM-EDX), while crystal disorders were followed with high-angle annular dark-field scanning transmission electron microscopy (HAADF-STEM). [12] Further details about growth, chemical procedures, and characterizations are given in the Methods. Crystallographic and composition variables are reported in the supporting information (SI, Table S1-S2, Figures S1-S3).

Figure 1(a) illustrates the crystal structure of a single layer of the pristine $NiPS_3$ (as processed from an SXRD), while Table 1 includes the related lattice parameters of the studied materials. The Table designates the monoclinic structure of the pristine and the intercalated $NiPS_3$ crystals [8] while emphasizing differences in their crystallographic parameters. The crystal axes "*a*", "*b*", and "*c*" of the intercalated crystals are expanded as compared to those of the pristine lattice, with a maximal unit cell volume expansion of 1.45 $Å^3$ in PY_50 (average expansion is ~1.11 $Å^3$). An intercalation process at temperatures higher than 75°C led to the deterioration of the sample. Moreover, crystallographic information given in the SI uncovered an occasional position exchange between the $Ni^{2+}$ ion and the $P_2$ pair, expressed schematically by the partially filled colors in Figure 1(a).

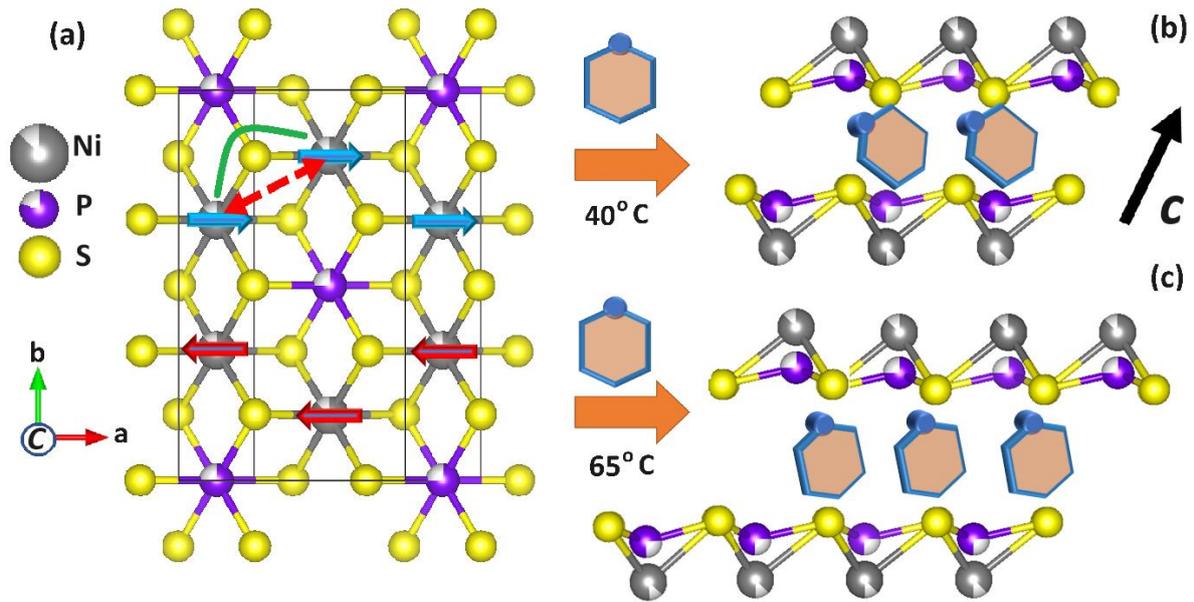

***Figure 1. Crystal structure analyses.*** *(a) A top view of the crystal structure obtained from an SXRD measurement at 100 K, representing the "ab" plane of a unit cell (black frame) of the pristine NiPS₃. The blue and red arrows designate the AFM spin alignment. The direct-exchange interaction is represented by a red-dotted double arrow, and the super-exchange interaction is shown by a green curve. (b, c) Pictorial representation of two vdW layers stacked along the c-axis and accommodating pyridine molecules (brown hexagons; blue circle – N atom) which were prepared at 40°C and 65°C, respectively. It is important to note that cross-sectional presentation (c) shows sliding between the stacked layers and a change of orientation of pyridine molecules concerning the "ab" plane.*

**Table 1.** SXRD lattice parameters of pristine and intercalated NiPS$_3$ single crystals at 100 K and variations at in-plane and out-of-plane Néel temperatures extracted from magnetization measurements. The standard uncertainties in lattice parameters are provided in the first brackets.

| Sample | NiPS$_3$ | PY_25 | PY_40 | PY_50 | PY_65 |
|---|---|---|---|---|---|
| Space group | C 1 2/m 1 | C 1 2/m 1 | C 1 2/m 1 | C 1 2/m 1 | C 1 2/m 1 |
| Lattice parameters (Å): *a, b, c* | 5.8032(3), 10.0574(5), 6.5978(5) | 5.8101(7), 10.0606(11), 6.8083(9) | 5.8096(7), 10.0596(10), 6.6093(9) | 5.8103(2), 10.0646(4), 6.6135(3) | 5.8099(3), 10.0637(4), 6.6142(3) |
| *α, β* (°) | 90, 107.019(7) | 90, 107.035(12) | 90, 107.057(12) | 90, 107.092(4) | 90, 107.114(5) |

| Δβ (°) =Pristine-(PY_X) | 0 | 0.016 | 0.038 | 0.073 | 0.095 |
| --- | --- | --- | --- | --- | --- |
| $V$(Å$^3$) | 368.22(4) | 369.33(8) | 369.27(8) | 369.67(3) | 369.60(3) |
| Δ $T_N\perp$(K) | 0 | 0.72 | 3.42 | 0.93 | 2.53 |
| Δ $T_N\|$(K) | 0 | 15.72 | 23.7 | 8.53 | 1.26 |

Beyond volume expansion, Table 1 underlines a significant change in the crystallographic parameter *β*, with a growth in the deviation (Δ*β* = Pristine-(PY_X)), along with the increase in the intercalation temperature (see Figure S4). In present study, Δ*β* has a more significant meaning than an expansion in "*c*", which is mainly discussed in previous publications [8, 32], proposing a nearly horizontal alignment of the pyridine molecule to the basal plane in the most loaded vdW gap with intercalants. However, those molecules gain an oblique angle upon sliding motion, as illustrated schematically in Figures 1(b) and (c). So, Δ*β* is used herein as a variable reference in analyzing other physical properties. It is essential to mention that the change in *β* is also accompanied by an expansion of the Ni-S and P-S bonds (see Table S2) and with internal disorders, like the antisite formation (see Figure S3), which will be further mentioned in the discussion below. [36-38].

HAADF-STEM images of cross-sections from pristine and intercalated (PY_65) crystals are shown in Figures 2(a) and 2(b, c), respectively. The related atomically resolved images (with color maps) are given in Figures 2(d-h). Figure 2(a) reveals a uniform stacking of the vdW layers along the pristine crystal's zone axis (0 1 0). A unit cell reconstruction at three different sections of the cross-section revealed the existence of a monoclinic unit cell with *β*~107.01°. The cross-sections of the intercalated crystal (Figures 2(b, c)) monitored at two different regions revealed inhomogeneity in the intercalants' distribution between adjacent layers (pay attention to different stains), and an average angle variable *β*~107.12° across different parts of the crystal. The above calculations from HAADF-STEM images indicate that the plausible origin of angular expansion of the unit cell of the SXRD (see Table 1 and Figures 1(b, c)) lies in the modification of stacking order upon non-uniform entry of intercalates [39, 40]. Complementary data in Figure S5 provides morphological evidence of the distortion of layer ordering after intercalation. The technique for extracting cross-sections of single crystals is provided in Figure S6.

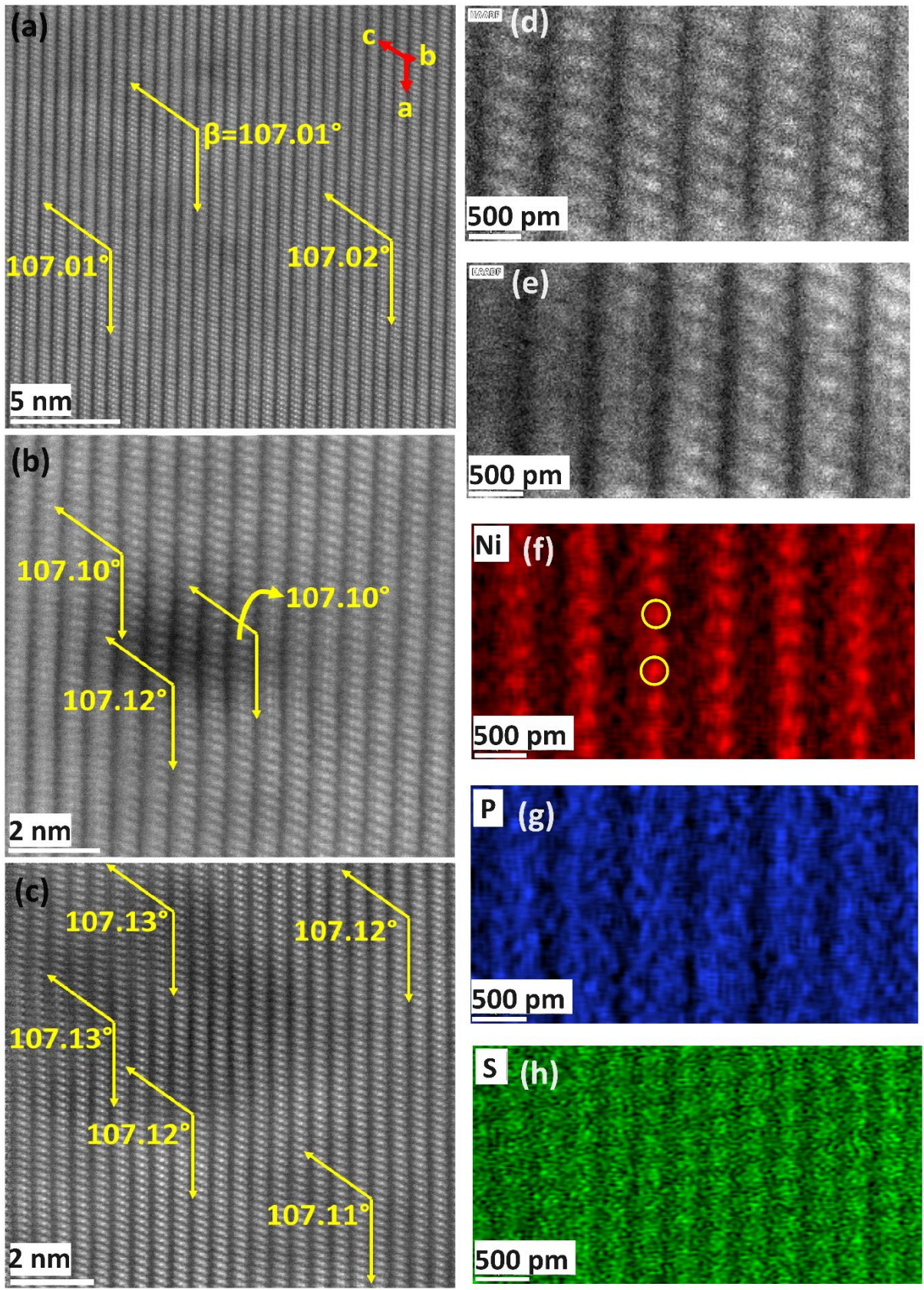

***Figure 2. Microstructural studies.*** *HAADF-STEM image of the cross-section of (a) a pristine NiPS$_3$ single crystal (b, c) PY_65, along zone axis (0 1 0). The crystallographic directions are*

*marked in (a). The calculated crystallographic angle β has been shown at different regions of the cross-sections. (d, e) Atomic resolution cross-section STEM image of NiPS$_3$ and PY_65, respectively. (f-h) HAADF images with atomic resolution of the elements in the NiPS$_3$ cross-section. The individual Ni atoms are highlighted in yellow circles.*

**Signatures of the intercalants.** The entry of pyridine molecules into the vdW gaps was investigated quantitatively by thermogravimetric analysis (TGA). The scheme for TGA measurements is provided in the Methods. Figure 3(a) shows a representative TGA curve of the intercalated crystal PY_50. The measurement shows a typical mass loss of ~2% at 115°C, characteristic of the boiling point of the pyridine molecules [32]. This deintercalation process counts to the presence of ~$1.52\times10^{17}$ pyridine molecules/mg of NiPS$_3$.

In addition, the Fourier transform infrared (FTIR) spectroscopy was used to detect the presence of prominent pyridine vibrations in intercalated and deintercalated NiPS$_3$ crystals. Figure 3(b) highlights the typical C-N stretching mode (around 1500 cm$^{-1}$) [41]. While this vibration mode is absent in the spectra of the pristine and deintercalated NiPS$_3$ samples, it is prominently pronounced in the spectra of the PY_50 and PY_65 crystals, with enhancement in the sample prepared at the highest temperature, hence containing a large number of molecules. The following panels in Figure 3 display Raman observations, to be mentioned in the following sections.

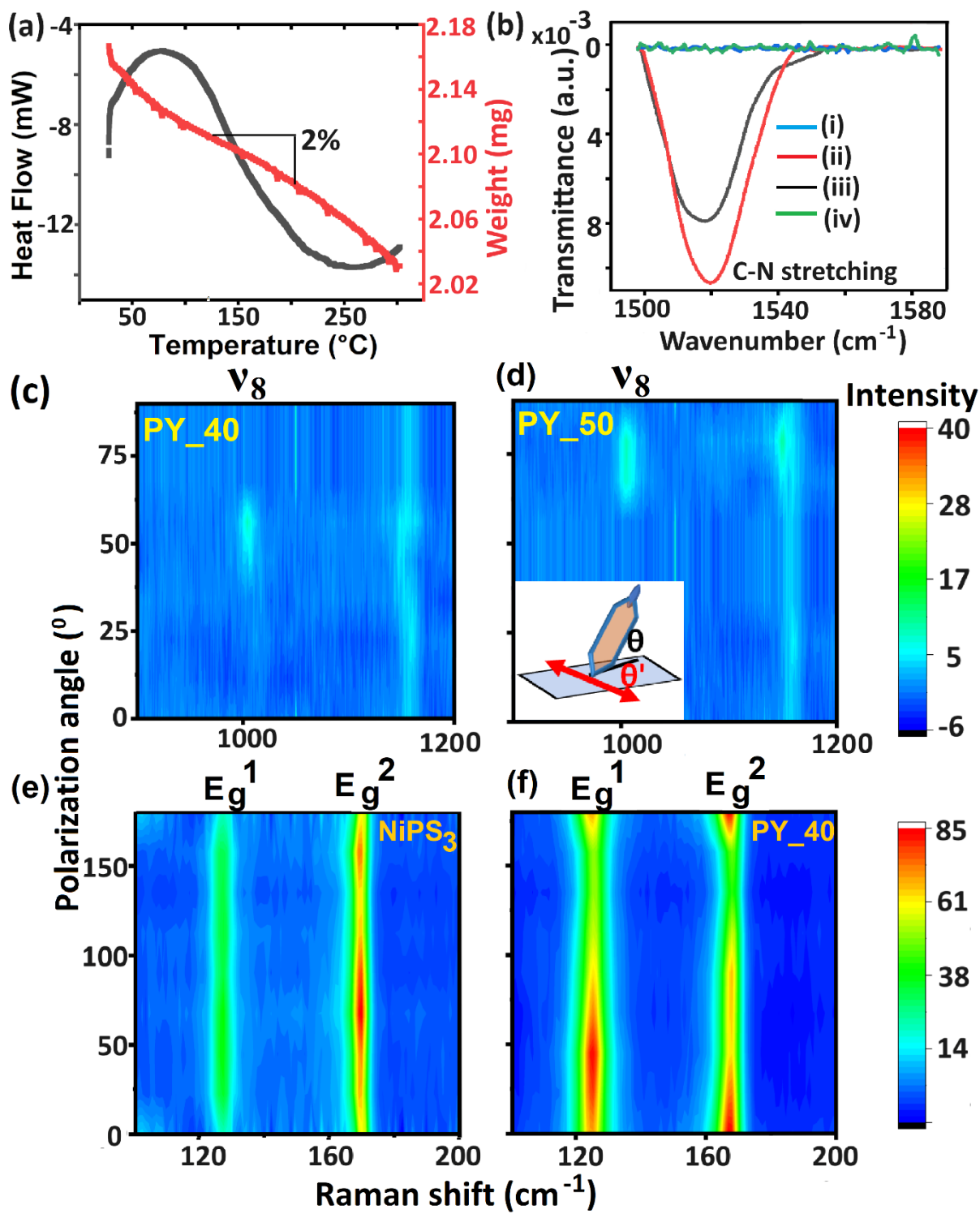

*Figure 3. Qualitative and quantitative identification of intercalates.* (a) A TGA measure of PY_50 under $N_2$ environment, showing a mass loss of 2%. (b) FTIR signal of C-N stretching mode of the pyridine intercalates for (i) pristine $NiPS_3$, (ii) PY_65, (iii) PY_50, and (iv) PY_50 after deintercalation (product collected after TGA). The curves in (i) and (iv) show no C-N vibrations. (c, d) Polarized Raman spectra of pyridine vibration at 1031 cm$^{-1}$ for PY_40 and PY_50, respectively. The inset of (d) shows a scheme for the Raman measurements, θ' being *the angle between the light electric vector and the projection of the pyridine dipole on the "ab" plane. θ is the angle between the pyridine dipole and the "ab" plane. (e, f) Polarization-dependent Raman spectra (recorded at room temperature) of pristine NiPS$_3$ and PY_40, respectively, related to the Ni atoms' vibration modes.*

The X-ray photoelectron spectroscopy (XPS) method was implemented to further verify the existence of intercalants and follow stoichiometric changes in the host crystal. Figure 4 displays XPS spectra of the pristine and one representative intercalated crystal (PY_40). The panels zoom into the various core levels of the elements, typified by their binding energies (B.E.). Figure 4(a) comprises Ni core-level spectrum, exposing the spin-orbit doublets 2P$_{3/2}$ and 2P$_{1/2}$ [42]. The peaks are deconvoluted with contributions arising from the presence of Ni$^{2+}$ and Ni$^{3+}$ ions (see pseudo-Voigt fits), and each is accompanied by weak satellites. The tendency of Ni for dual oxidation appears in Ni-based compounds with vacancies and defects (see Table S1). [43-45]. Figure 4(d) depicts the XPS spectrum of intercalated crystal PY_40 around the Ni doublet 2P$_{3/2}$ and 2P$_{1/2}$ region, revealing shifts to higher binding energies after intercalation. Figures 4(b, c) and 4(e, f) zoom into the P 2p and S 2p levels, exposing shifts to higher energies upon intercalation. Finally, Figures 4(g-i) illustrate the levels of N 1s where the deconvolution exposes contribution from C-N and -N-H$^+$ components, associated with pyridine and pyridinium ions, respectively [32, 46]. The pyridinium signal is generated *in-situ* via condensation of two pyridine molecules followed by the formation of bipyridyl ion and the releasing of two electrons to the NiPS$_3$ host [32, 46]. Table 2 summarizes the samples' composition (as chemical formula) before and after intercalation. The tables uncover a strong correlation between S-deficiency or intercalant-host charge donation with the occurrence of dual oxidation states for the metals – pronounced by the increase of the Ni$^{3+}$/Ni$^{2+}$ ratio with the increase of intercalants' concentration and the *Δβ* outcome. Supposedly, such a correlation is stimulated by the need for charge compensation. Furthermore, Table S3 reflects an increase of B.E. for Ni, P, and S-core levels and an extension in their mutual bond length [47, 48]. In contrast, there is a reduction of B.E. for the pyridine/pyridinium levels, in correlation with their bond-length shrinking [47, 48]. Redistribution of charges between intercalants and host and the angular distortion of the lattice (without significant modification in inter-layer gaps) is the reason for the changes in bond lengths. Additional XPS observations related to the PY_65 sample are provided in Figure S7.

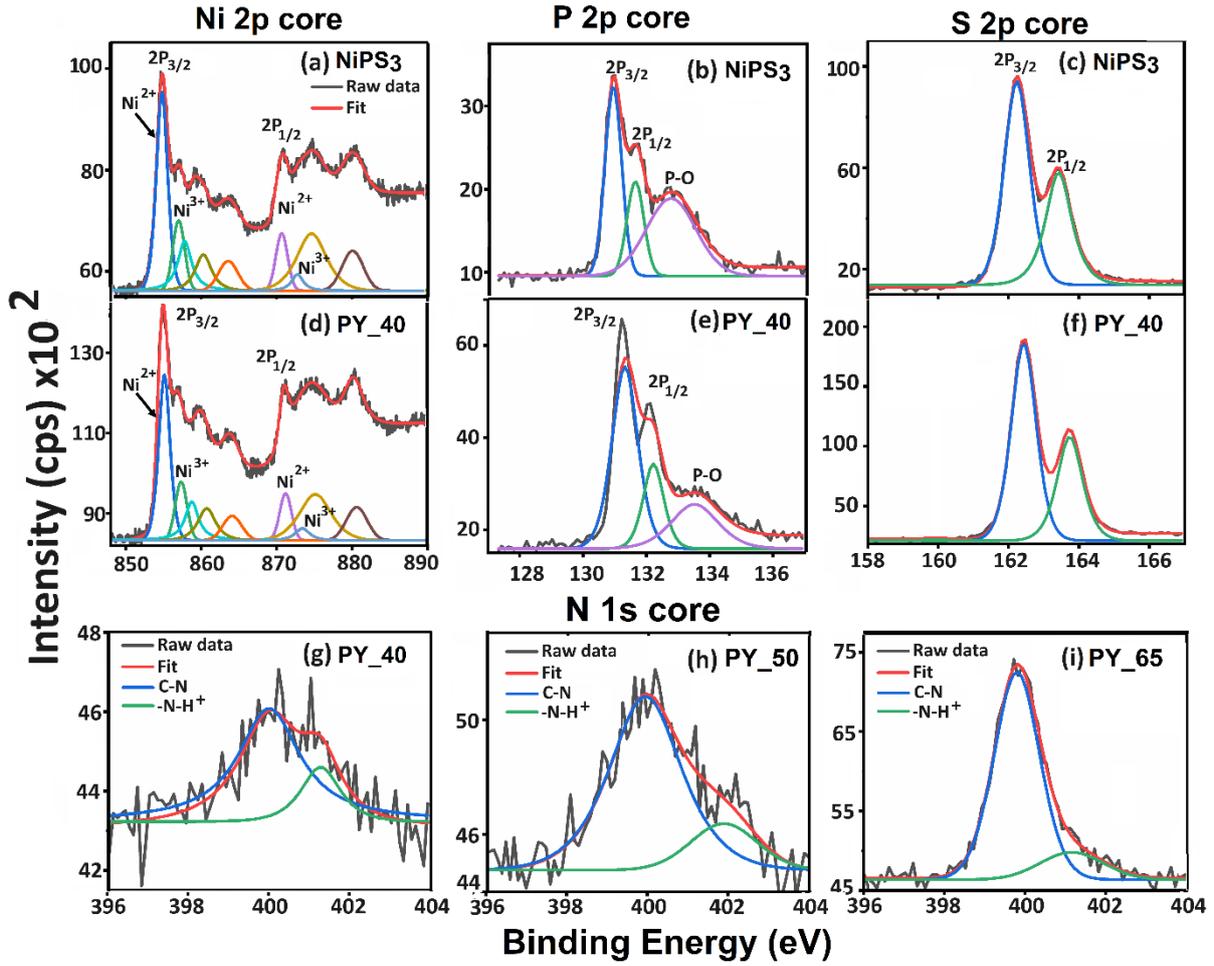

*Figure 4. Surface electronic state analyses.* *(a, d) Ni 2p core-level spectra deconvoluted into spin-orbit doublet states of $2P_{3/2}$ and $2P_{1/2}$ for pristine $NiPS_3$ and PY_40, respectively. (b, e) P 2p core-level spectra for pristine $NiPS_3$ and PY_40, with spin-orbit doublet states of $2P_{3/2}$ and $2P_{1/2}$ and additional P-O signal due to surface chemisorbed oxygen (c, f) S 2p core-level spectra, with spin-orbit doublet states of $2P_{3/2}$ and $2P_{1/2}$, for pristine $NiPS_3$ and PY_40, respectively. (g-i) N 1s core-level spectra, indicative of pyridinium, in PY_40, PY_50, and PY_65, respectively.*

**Table 2.** Quantitative estimation (in arbitrary units) of surface elemental composition from XPS. $e^-/\mu^3$ means the number of electrons transferred to the host (per $\mu^3$) due to the formation of pyridinium ions.

| Sample | $Ni^{3+}:Ni^{2+}$ | S | P | $NH^+$ | $e^-/\mu^3$ ($\times 10^{19}$) | Chemical formula |
|---|---|---|---|---|---|---|
| $NiPS_3$ | 0.081 | 2.892 | 0.997 | — | — | $Ni_{0.925}^{2+}Ni_{0.075}^{3+}P_{0.997}^{4+}S_{2.892}^{2-}$ |
| PY_25 | 0.086 | 2.883 | 0.972 | 0.008 | 9.6 | $Ni_{0.920}^{2+}Ni_{0.080}^{3+}P_{0.972}^{4+}S_{2.883}^{2-}$ |
| PY_40 | 0.091 | 2.878 | 0.967 | 0.010 | 12.0 | $Ni_{0.916}^{2+}Ni_{0.084}^{3+}P_{0.967}^{4+}S_{2.878}^{2-}$ |
| PY_50 | 0.097 | 2.869 | 0.959 | 0.012 | 14.4 | $Ni_{0.911}^{2+}Ni_{0.089}^{3+}P_{0.959}^{4+}S_{2.869}^{2-}$ |
| PY_65 | 0.102 | 2.865 | 0.948 | 0.015 | 18.0 | $Ni_{0.907}^{2+}Ni_{0.093}^{3+}P_{0.948}^{4+}S_{2.865}^{2-}$ |

The orientation of the pyridine molecules within the vdW gap was determined from an optical linear polarization of a Raman mode [49-55]. Pyridine molecule has been reported to belong to the $C_{2v}$ point group, with 27 Raman active modes, out of which the modes with $A_1$ symmetry (where the moment of inertia lies along the N-$C_4$ figure axis) is polarized [49, 50]. Figures 3(c, d) represent the polarization-dependent Raman spectra of two samples, PY_40 (c) and PY_50 (d), as color plots. The orientation is described as the tilt angle between the pyridine dipole axis (with a N atom at the head) and the "*ab*" plane of the host lattice (see the inset of Figure 3(d)). The procedure to estimate the probable orientation of the pyridine molecule to the "*ab*" plane has been given in the Methods. Figures 3(c, d) contain two resolved modes, one of which is the polarized $v_8$ mode ($A_1$ symmetry) at ~1031 cm$^{-1}$ (corresponds to in-plane ring bending) [54]) and shows a pronounced polarization at certain angles (45-60° for PY_40 and 60-80° for PY_50). Maximal polarization appears when the optical dipole moment coincides with the pyridine dipole direction, thus revealing the molecular orientation (θ in the inset of Figure 3(d)). Interestingly, the full set of data (not shown here) showed an approach to a normal alignment of the pyridine to the "*ab*" plane with the increase of the *Δβ* (viz., the largest sliding of adjacent layers upon intercalation, as shown schematically in Figure 1(c)). It is important to mention that the Raman polarization was pronounced only at certain regions of the sample, thus confirming the non-uniform distribution of intercalants as found in the HAADF-STEM images (Figures 2(b, c)). Figure S7 shows the pyridine orientation of the PY_65 sample as complementary data. The impact of intercalant on imparting polarity to the $NiPS_3$ host was identified primarily by polarization study of Raman active $E_g^1$ (corresponds to Ni-S stretching) and $E_g^2$ (Ni-S bending) modes [51, 52]. Figures 3(e, f) highlight the polarized Raman spectra for pristine $NiPS_3$ and PY_40, revealing increased (doubled for PY_40) degree of linear polarization of the Ni-S bond stretching. This can be attributed to the additional charge transfer to Ni by pyridinium, impacting its bonding environment. Additional results on polarized Raman spectra are given in Figure S8. Figure S9 alternatively validates the generation of polarization, *via* capacitive measurements, in the $NiPS_3$ samples by virtue of the intercalant [56].

**Magnetic Anisotropy.** As discussed in the introduction, the NiPS$_3$ host crystals possess anti-ferromagnetism (AFM) with a zigzag configuration along the "*a*" crystal axis. This study exposed the influence of the pyridine intercalation on the magnetic anisotropy of the host lattice by following the magnetic susceptibility ($\chi$) using the superconducting quantum interference device (SQUID) method. Figures 5(a-c) display field-cooled magnetic susceptibility versus temperature measurement plots of pristine (a), PY_40 (b) and PY_65 (c) intercalated crystals. The samples experienced the influence of an external static magnetic field (H) of 5000 Oe, applied parallel ($\parallel$, blue) or perpendicular ($\perp$, red) to the "*ab*" plane (see legends) [22, 28]. The curves display typical trends of anti-ferromagnetism, with breaking points associated with AFM-paramagnetic phase transition (marked next to each curve), and known as the Nèel temperature, $T_N$. The values of $T_N$ were extracted from the singularity in respective $d\chi/dT$ versus temperature curves, provided in Figure S10. Figure 5(f) represents plots of the extracted $T_N$ values of various samples versus their related $\Delta\beta$ parameters. The error bars represent a standard deviation among three different sets of data. The plots designated contrast changes between $T_N(\parallel)$ and $T_N(\perp)$, proposing an optimal point (associated with composition PY_40) in which *additional magnetic vector was developed along the trigonal axis of the host lattice, hence altering the anisotropy in the host lattice*. The gradual increment in out-of-plane susceptibility ($\chi_\perp$) in Figures 5(b, c) indicates at the generation of this out-of-plane magnetic order (compared to Figure 5(a)). This optimal point degrades upon further loading of intercalants, supposedly due to a larger disorder. However, the disorder is reversible *via* a thermal treatment in the vacuum at 200°C, a point in which a full de-intercalation takes place and the NiPS$_3$ host lattice gains the original in-plane magnetization with a 1% deviation from the initial unintercalated state (see Figure S11). Additional susceptibility-temperature curves are provided in Figure S11. The above trend in $T_N(\parallel)$ and $T_N(\perp)$ was further confirmed using polarized Raman spectroscopy at different temperatures, where the discontinuity in doubly degenerate spectra of $E_g^2$ mode indicates the phase transition point (onset of spin-phonon coupling) [5, 57]. Figures 5(d, e) show the concerned Raman spectra involving the $E_g^2$ mode in $z(y\bar{x})\bar{z}$ configuration (see the inset of (e)), for pristine NiPS$_3$ and PY_40, indicative of an increase in $T_N(\parallel)$ with intercalation. The phase transition temperatures are marked by yellow squares in Figure 5(f). Additional results for the $z(yx)\bar{z}$ configuration (indicative of $T_N(\perp)$) and the raw data are supplied in Figure S12.

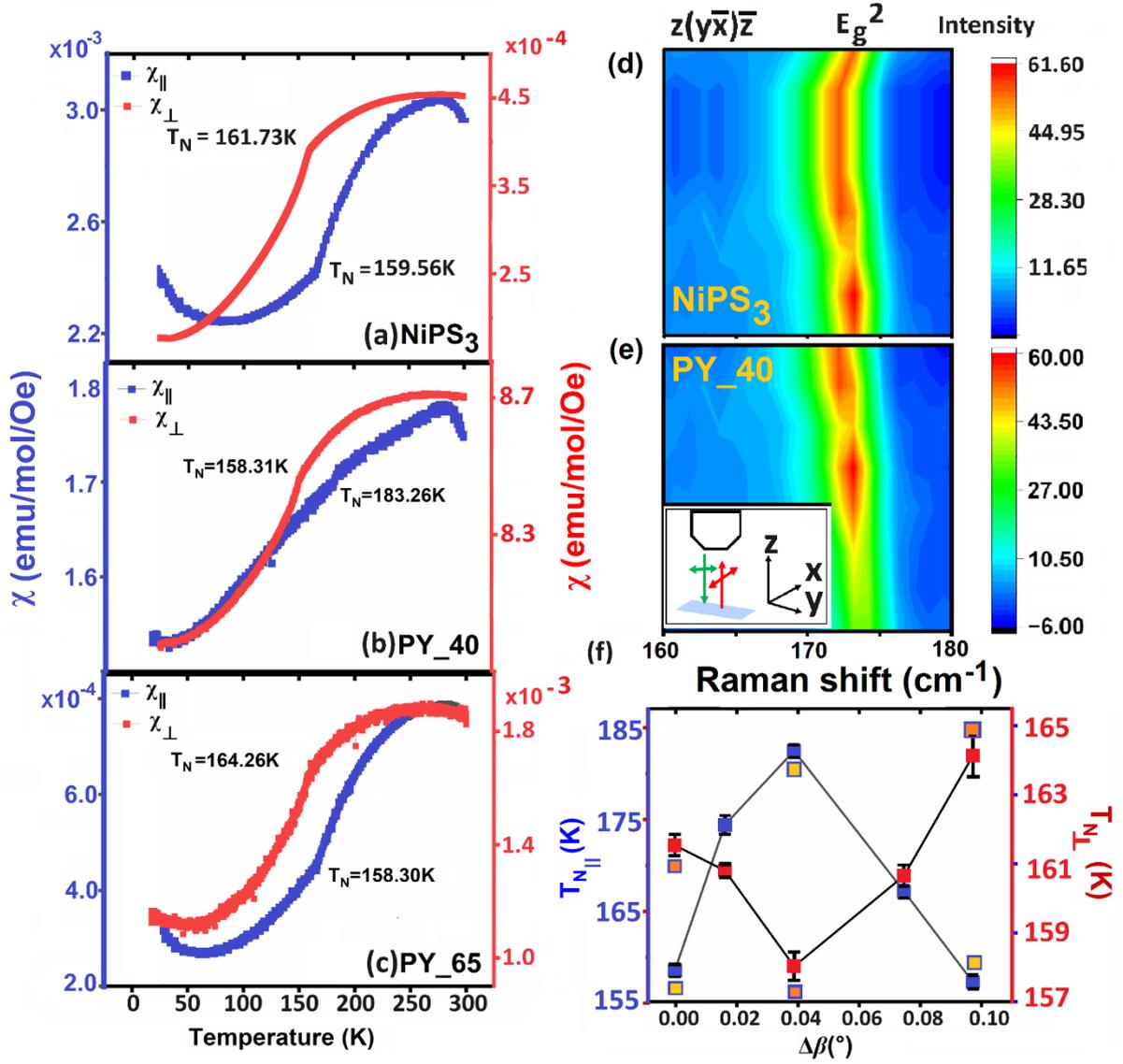

*Figure 5. Magnetic property measurements.* *(a-c) Magnetic susceptibility vs. Temperature curves for pristine and intercalated NiPS$_3$ crystals in the field cooling (FC) mode under an external field (H) of 5000 Oe. $T_N$ represents the Néel temperature of the corresponding in-plane (H ∥ to ab plane) and out-of-plane (H ⊥ to ab plane) configurations. (d, e) Temperature-dependent variation of the $E_g^2$ Raman mode for pristine NiPS$_3$ sample and PY_40 in $z(y\bar{x})\bar{z}$ configuration. Inset of (e) shows the schematic of the geometry of measurement used, involving the microscope head, incident (green) and emitted(red) light and corresponds to the $z(yx)\bar{z}$ notation. (f) An overview of the changes in in-plane and out-of-plane magnetic phase transition temperatures with change in Δβ. The yellow boxes represent the $T_N(∥)$ phase transition values and orange boxes the $T_N(⊥)$ values obtained from temperature-dependent polarized Raman spectroscopy. The error bars (black color) represent the standard deviation over 3 different sets of data collected with 3 different sample sets.*

**Thermal Transport.** The above dynamics in intra and inter-layer spin properties bring into consideration the role of inter-layer thermal transport that can be anticipated to affect spin ordering by establishing thermal contacts between the otherwise weakly coupled vdW layers [58]. Thermal transport measurements were carried out to uncover the role of intercalated molecules in enhancing inter-layer connection. The experiment included monitoring the specific heat, that, by further modeling (using Fick's Law, see Methods), [59,60] provided the heat diffusion coefficients. Figure 6(a) illustrates the experimental setup, labeling heat flow by the graded color arrows (dark color refers to the warmer side). The bottom substrate resistance is marked by R. Figure 6(b) displays plots of the measured specific heat versus temperature for the samples under investigation. The deflection points in the curves designate magnetic phase transitions at $T_N$, which are correlated with the magnetic susceptibility measurements. Analysis of the heat capacity curves (see Methods) revealed the heat diffusion coefficient of the in-plane and out-of-plane components (referred to as $D_X+D_Y$ and $D_Z$, respectively). Figure 6(c) displays the dependence of the mentioned coefficients on $\Delta\beta$, exposing an increase in the out-of-plane heat diffusion as $\Delta\beta$ increases, *viz.,* increasing the intercalants' concentration and creating a wider channeling between adjacent layers. Finally, Figure 6(d) displays plots of thermal conductivity versus temperature of the various samples as in panel (b), which showed a tendency for enhancement of heat transport across the intercalants (z-direction) (with an exception for the most compacted vdW gap due to disorder in the latter). Preliminary electrical conductivity measurements revealed enhanced conduction in intercalated samples (see Figure S8), requiring further studies in this way.

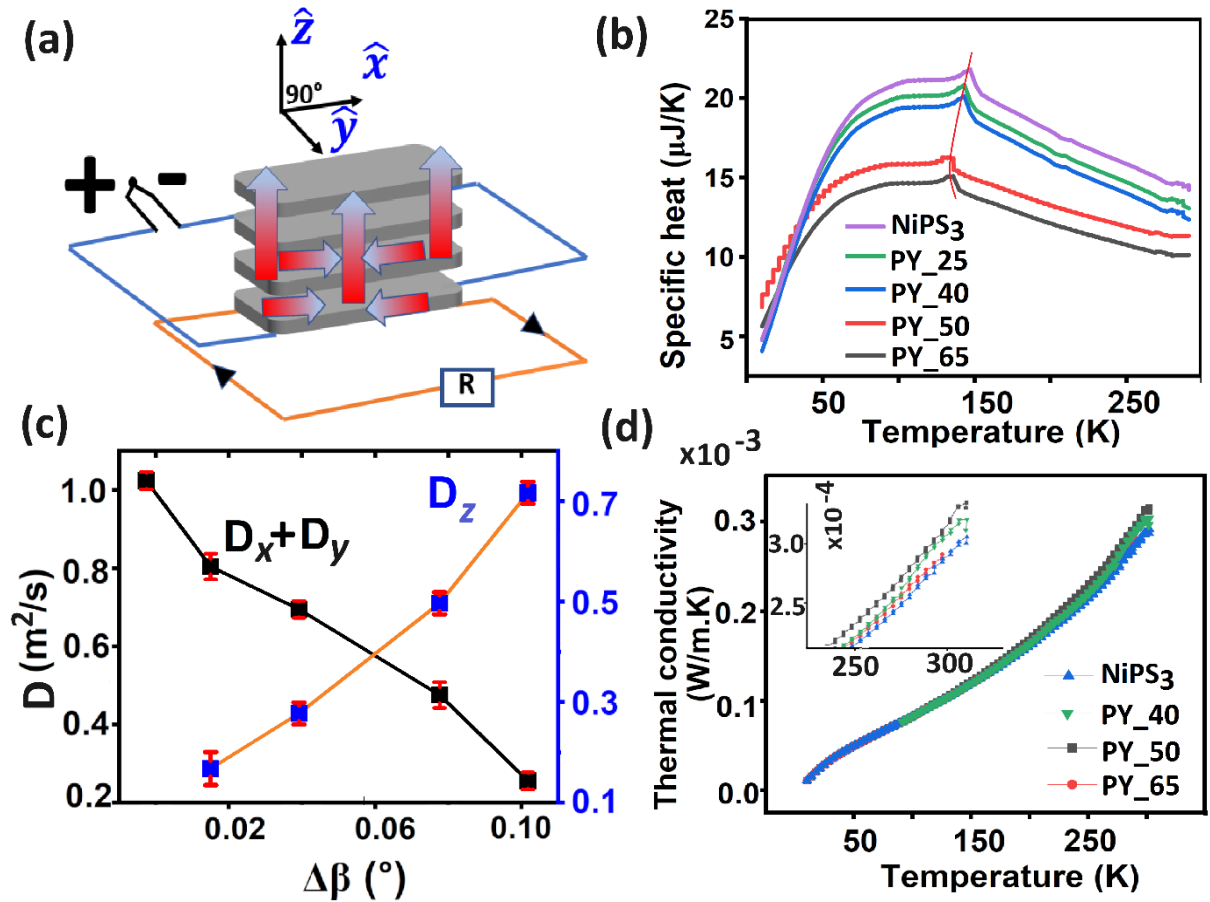

*Figure 6. Thermal property measurements.* *(a) Schematic of thermal transport measurement in layered material like NiPS$_3$. "R" represents resistance for a 2D heater placed below the sample consisting of layers depicted in gray. The thermal gradient is highlighted as arrows, with a color gradient representing more heat at lower levels due to proximity to the heat source. (b) Variation in specific heat with temperature for pristine and intercalated samples. The shift in phase change temperatures is highlighted by the red line. (c) Variation in in-plane and out-of-plane thermal diffusivities with the increase in Δβ. The error bars represent standard deviation in modeling the thermal diffusivity using Fick's second law differential equation. (d) Thermal conductivity curves for pristine and intercalated samples. Inset shows the variations on a bigger scale for better representation.*

**Modeling and an overall discussion**

The research described above characterized the structure, composition, and components' orientation of pyridine intercalated NiPS$_3$ vdW magnetic materials. Furthermore, the study investigated the changes in magnetic and thermal properties emanating from the intercalation process. The discussion here proposes the most plausible mechanism responsible for the variations in anisotropy in magnetism and heat transport. Previous publications involving metal /organic ions intercalation into $NiPS_3$ *via* electrochemical route proposed the emergence of ferrimagnetism (FIM), bestowed either by the formation of $Ni^0$ species, thus, random reduction of metal ions across the honeycomb arrangement [20, 21]. Other studies suggested a selective occupation of a singular Ni zigzag by a donating electron from the intercalants, hence forming unequal magnetic columns and consequent formation of ferrimagnetism. [22, 61]. The observations of this study solely exclude the formation of FIM due to the absence of its markers (e.g., net FM-like magnetism) in the magnetic susceptibility measurements. Also, $Ni^0$ did not appear in the XPS observations.

Alternatively, we support a model named hereon as "double spin-exchange" as the mechanism that explains the emergence of the out-of-plane component in the magnetic and thermal measurements. The model has been proposed before [23, 24] to address the influence of intercalant-to-host charge injection and the main reason for property variations. The principles of the models are explained schematically in Figure 7.

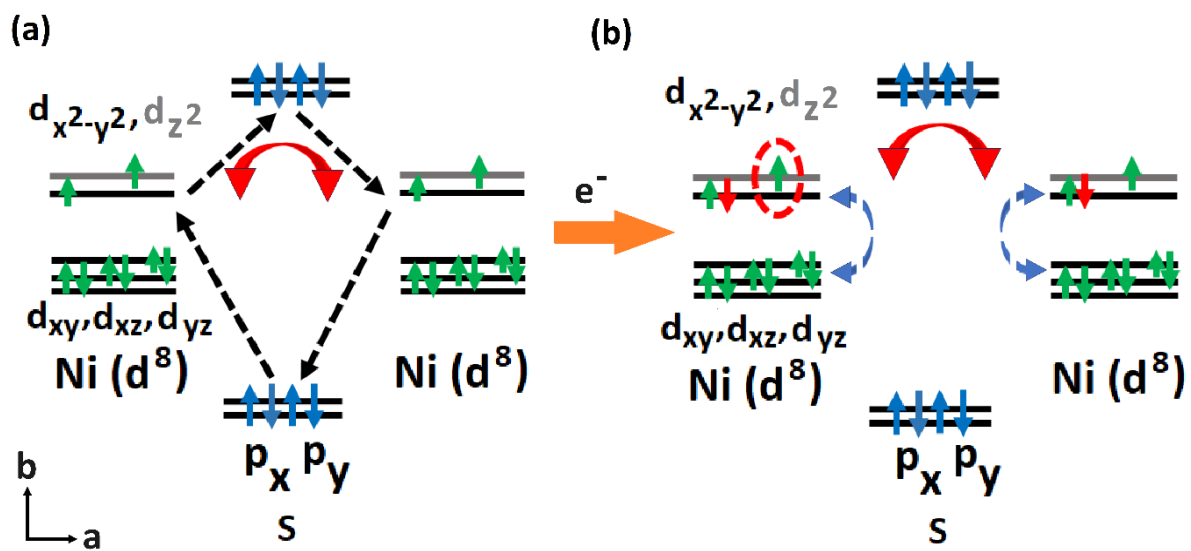

*Figure 7. Spin mechanism (a) Proposed exchange mechanism in pristine and intercalated $NiPS_3$ based on atomic arrangement (top view) depicted in Figure 1a. The most probable scenario of super-exchange in pristine $NiPS_3$ via Ni-S-Ni is represented by black dashed arrows. The direct exchange is represented by a double-red arrow. (b) The red colored spins are representative of the contribution from electron-donor intercalates. The blue arrow indicates the Hund's coupling. The unpaired electron in $d_z^2$ state (encircled in red) becomes deterministic in controlling the magnetic order. This reveals the onset of double-exchange interaction.*

Figure 7 demonstrates the possible routes for magnetic coupling between two adjacent metal ions (see Figure 1(a)), a process playing a role in controlling the long-range AFM magnetism. Each metal ion comprises eight spins in its d-orbits, spread according to the crystal-field states, $t_{2g}$ and $e_g$ ($t_{2g}$: $d_{xz}$, $d_{xy}$, $d_{yz}$; $e_g$: $d_{x^2-y^2}$, $d_{z^2}$). According to Hund's rule, the $t_{2g}$ is fully occupied, while $e_g$ is half field. In the pristine sample, the next neighbor spin-exchange interaction involves a super-exchange *via* two S(p) states (based on hole hopping between those states) (see Figure 7(a)) [23] [24]. The super-exchange processes are marked by the black dashed lines.

Upon intercalation, the donated charge by the intercalant would occupy a half-filled state, immediately un-balancing the network of super-exchange coupling among neighboring metals. The suppression of the super-exchange opens another route for coupling, the "direct spin-exchange", shown by the red curved double arrow. [20, 21, 29, 61, 62]. The donated electron preferably occupied the $d_{x^2-y^2}$ state as seen by the red arrow in Figure 7(b). In addition, the nearly full d-state occupation stimulates spin distribution among the d-states (the so-called Hund's exchange coupling [23, 29, 31]). This spin-spin coupling increases the $T_N\|$ values, as shown in Figure 5(f). The $d_{x^2-y^2}$ state is no longer available for direct exchange, and therefore, it is only the $d_z^2$ electron that dominates it, hence capable of playing a decisive role in determining a local or a global magnetic ordering (see Figure 7(b)). It is noteworthy that the $d_z^2$ orbitals are dumbbell-shaped, perpendicular to their donut-like ring, and thus, bestows a z-component to the magnetization. Indeed, the post-inflection point in Figure 5(f) showed a depletion in the $T_N\|$ values due to reduction in Hund's coupling and enhanced disorder by thermal transport, reducing in-plane coupling strengths. The highest concentration of intercalants, viz., enhancement of carriers' donation, induces a prominent out-of-plane magnetic component. It has been seen by an increase of the $\chi_\perp$ value in Figure 5(c). However, further filling of the $e_g$ states would reach saturation that blocks any further increase of normal magnetization. Investigations into the intercalation of other organic/inorganic molecules in different members of the $MPX_3$ family, and similar studies on them are underway.

**Summary**

NiPS$_3$ single crystals were intercalated with pyridine molecules under various thermal conditions. The dynamic variation in internal electronic structure, while accommodating the intercalates, has been studied in terms of global and local structure modifications. Global structure studies reveal the gradual expansion of the unit cell with a sequential increase in the crystallographic angle *β* (as the intercalation temperature increases), accompanied by an expansion of the Ni-S and P-S bonds. Local structure investigations using X-ray photoelectron and Raman spectroscopies depict a significant and selective role of pyridine intercalates in initiating charge transfer mechanisms with host $NiPS_3$ and their relative orientations. As evident from magnetic measurements, the inclusion of pyridine dipoles obliquely between the vdW layers leads to stronger out-of-plane components of magnetic moments in the otherwise XY antiferromagnetic system. Contrary to the out-of-plane Néel temperature variations, the variation of in-plane Néel temperatures with an increase in $\Delta\beta$ shows a reverse nature. The theory of double spin-exchange interaction has been used to explain the magnetization results. The above pattern of magnetic orientation is supported by the variations in 3D thermal diffusivity through the connected vdW layers, depicting the interplay of thermal disturbances and spin exchange interactions with material electronic structure in modulating the magnetic anisotropy. The above process is reversible, following the removal of the pyridine intercalates *via* thermal deintercalation. This introduces the possibility of using intercalation as a method to achieve controlled and preferential magnetic ordering in magnetically anisotropic lamellar systems. Modifying the orientation of pyridine molecules further by external electric fields can make the switching process more precise and regulated. Layer-dependent magneto-optical studies of intercalated systems can provide further information on the role of various quantum entities in controlling the above spin phenomena under different conditions of intercalation.

**Methods**

**1. (a) Synthesis of $NiPS_3$ single crystals**

Single crystals of $NiPS_3$ were prepared by CVT method, where Ni (Aldrich, 99.99%), P (red, Riedel-de Haën, 97%), and S (Aldrich, 99.98%) powders were taken in 1:1:3 ratio with 5% extra sulfur as transporting agent and mixed thoroughly [63]. The above mixture was transferred to a clean and dry quartz ampoule and sealed under a vacuum of ~$10^{-5}$ Torr. The sealed ampoule was then put into a two-zone furnace with the hot zone at 750°C and the cold zone at 690 °C, over one week. Big crystals of approximate size 0.2 mm * 0.3 mm * 1 μ were collected from the product zone. The size and thickness of the crystals were measured using an optical microscope and Atomic Force Microscopy (AFM) (see Figure S6).

**(b) Synthesis of intercalated $NiPS_3$**

For pyridine intercalation, about 5 mg of the $NiPS_3$ sample (single crystals) was taken into a round bottom flask together with 0.5 ml of anhydrous pyridine (Fisher Chemical, >=99.5% with 0.02% water). The flask was then sealed with cork and grease and flushed with dry $N_2$ for 2 minutes. The flask was then placed on a heater at different temperatures (room temperature, 40°C, 50°C, 65°C, and 75°C) for 4 days and then cooled down. The crystals were then washed with ethanol, dried, and stored in vacuum. TGA of the prepared samples was performed by heating the crystals in an inert ($N_2$) environment using a Mettler Toledo TGA instrument (see Figure 3(a)), at a heating rate of 1°C/min and cooled at 0.7°C/min. The onset of mass loss was estimated by calculating the derivative of the TGA curve with respect to temperature. Consequently, the deintercalation was carried out by heating the intercalated crystals at 200 °C under vacuum for 2 days and brought down to room temperature with a cooling rate of 0.7°C/min.

## 2. Crystal structure and microstructure analyses

Crystal structure of pristine and intercalated $NiPS_3$ were determined in their magnetic AFM phase (100K) in a Rigaku Synergy S diffractometer with hybrid pixel CdTe Dectris 3R 300K detector and Oxford cryostream plus cooling system, using Mo $K_\alpha$ radiation (see Figure 1(a) and Figure S3) [64]. Numerical absorption correction was based on Gaussian integration over a multifaceted crystal model. Empirical absorption correction was done employing spherical harmonics [65]. Data integration was carried out using the CrysAlisPro 1.171.43.104a software [66], and the single crystal structure was solved using Olex 2 software [67]. Initial identification of the lattice by indexing diffraction peaks gave a monoclinic unit cell with a "C 1 2/m 1" space group (see Table 1). Structure solution in pristine $NiPS_3$ was a challenge owing to the small ionic radius of $Ni^{2+}$ (~0.83 Å in high-spin octahedral geometry) as compared to $Mn^{2+}$ (~0.97 Å) or $Fe^{2+}$ (~0.92 Å) of the $MPX_3$ family of 2D antiferromagnets [36]. This made the determination of the positions of Ni and P difficult. On replacing some Ni with P and vice versa, the refinement iterations led to a good fit (see Figure 1(a) and Figure S3).

SEM images were captured using an Ultra Plus Zeiss FEG-SEM at an accelerating voltage of 15-16 kV with an SE2 (Everhart–Thornley) detector (see Figure S5). EDX spectra and elemental mappings were obtained using an X-Max Silicon Drift Detector under the same beam conditions as the SEM images (see Figures S1 and S2). Cross-sections of the single crystals were collected in a Thermo Fisher Helios 5 Plasma Focused Ion Beam (PFIB) Workstation (see Figure S6). The above cross-section samples were further loaded into the TEM machine for STEM measurements. The HAADF-STEM micrographs and the EDX maps were acquired using a monochromatic and double-corrected Titan Themis G2 60-300 (FEI/ Thermo Fisher) operated at 200KeV and equipped with a DualX detector (Bruker) (see Figure 2). The EDX maps were analyzed using the Velox software (FEI/Thermo Fisher) (see Figure 2). The lattice parameter ($\beta$) was calculated from the STEM images using the CrysTBox software (see Figure 2) [68].

## 3. Spectroscopic studies

Identification of surface electronic states was carried out by XPS in Versaprobe III–PHI Instrument (PHI, USA) using monochromatic Aluminum K-alpha X-ray of photon energy 1486.7 eV as a source at a pass energy of 55 eV and step size of 0.1 eV (see Figure 4, Tables 2 and S3, and Figure S7). Background correction of core-level spectra was done by employing the Shirley algorithm, [69] and the chemically distinct species of Ni, P, S, and N were resolved by a non-linear least-squares fitting method using the software KoIXPD [70]. The core-level binding energies of individual elements were scaled with the adventitious carbon peak with a binding energy of 284.8 eV. To investigate the possible existing oxidation states of Ni, P, S, and N atoms, the area under Ni 2p, P 2p, S 2p, and N 1s core-level signals were deconvoluted employing a combination of Gaussian and Lorentzian profile functions (see Figure 4). The quantitative analyses of the existence of various oxidation states of Ni, P, S, and N were obtained by scaling the area under XPS curves by the photoionization cross-sectional values of 2p orbitals of Ni, P, S, and 1s orbital of N under Al $K_\alpha$ radiation respectively (see Table 2).

FTIR spectroscopy was performed in the range of 800-3600 cm$^{-1}$ using Thermo Scientific Nicolet iS50 instrument. Polarization-dependent Raman measurements were performed using Witec Alpha 300 micro-Raman instrument, employing a 532 nm laser. The optical pathway was initially calibrated with a standard Si/SiO$_2$ sample using 100X (Numerical Aperture 0.9) objective and then used to measure the single crystal mounted on a Si/SiO$_2$ wafer. A laser power of 5mW was used with grating 1200/500, an integration time of 5 s, and 20 accumulations to get one spectrum. Spectra were obtained at 3 different spots of 1 µ diameter. Polarization measurements were done in the same set-up with a polarizer varying from 0° to 90° at intervals of 5° and an analyzer kept at 0°.

The schematic to understand the methodology for estimating the degree of inclination of pyridine molecules from polarized Raman is provided in the inset of Figure 3(d). The laser source, being plane polarized, had different orientations of its electric field $\vec{E}$ component, parallel to the "*ab*" plane. The signal from the interaction of incident electric field $\vec{E}$ with the dipole moment of pyridine molecule $\vec{P}$ leads to potential energy U=-$\vec{P}$. $\vec{E}$=-P.E. Cos(θ+θ'), where θ is the angle between pyridine dipole and "*ab*" plane, θ' is the angle between $\vec{E}$ and projection of the dipole on the "*ab*" plane. From the configuration of the laser and sample in Raman measurements (see the inset of Figure 3(d)), the formula stands as U=-$\vec{P}$. $\vec{E}$ = -P.E. Cos(θ+θ') = P.E. [(Sin θ).(Sin θ') - (Cos θ).(Cos θ')]. For identifying the polarization of the Raman signal from pyridine, the polarizer (hence $\vec{E}$) was rotated from θ'=0° to 90°.

## 4. Magnetic and transport property measurements

Magnetization measurements were carried out in a Quantum Design Magnetic Property Measurement instrument, in both in-plane and out-of-plane configurations (see Figure 5). For in-plane mode, the single crystals (approximately 3.5 mg) were attached to a quartz holder with GE varnish and for out-of-plane mode, the crystals were held on a quartz bead attached to a brass holder. DC Magnetic moment vs. temperature studies were carried out under an external field of 5000 Oe in both Field Cooling (FC) and Zero Field Cooling (ZFC) modes. The samples were cooled in zero field while being measured, with a cooling rate of 2 K/min. Then, at 1.8 K, the magnetic field was raised to 5000 Oe and the samples were measured while heating to 300K and then cooling to 300K at the same 2 K/min rate. The quartz holder has a strong diamagnetic signal of its own. The final magnetization values have been obtained after subtracting the signal from the holder. The heat capacity measurements were performed with the Quantum Design Dynacool Heat Capacity module. The thin flat samples (approximately 5.6 mg) were placed on the sample mounting platform, nearly covering it all. The heater and thermometer are placed on the bottom part of the platform, hence applying heat from below the sample, in a reasonably uniform manner.

The 3D heat transfer model was developed based on Fick's second law [59,60], where the change in concentration gradient of heat flow with time can be equated as:

$$\frac{\partial C}{\partial t}=D.\nabla^2 C \dots\dots\dots\dots\dots\dots\dots(4)$$

$$\text{or } \frac{\partial C}{\partial t}=D_x.\nabla_x^2 C + D_y.\nabla_y^2 C + D_z.\nabla_z^2 C \dots\dots\dots\dots(5)$$

where C(x,t) is the concentration of heat flow as a function of space and time, D is the thermal diffusivity in m$^2$/s and $\nabla^2 C$ is the term for the spatial distribution of heat in 3D. On fitting the $\frac{\partial C}{\partial t}$ curve using three Laplacian terms, we get three components that correspond to thermal diffusivity along $\hat{x}$, $\hat{y}$ and $\hat{z}$ directions, respectively. If we consider the plane of the vdW layer to be the base and the thermal gradient along $\hat{Z}$, then the $D_x$ and $D_y$ terms provide the in-plane heat transfer whereas the $D_z$ term provides the out-of-plane thermal diffusivity (see Figures 6(a) and 6(c)). The fitting parameters for the non-linear curve fitting are provided in Table S4. Thermal conductivity was measured in a DTC 300 set up from TA instruments from 10K to 300K (see Figure 6(d)).

**Acknowledgement**

E. L. and D. N. thank the support from the Deutsch-Israel Program (DIP, Project No. NA1223/2-1). E.L. thanks the Israel Science Foundation (ISF, Project No. 2528/19). The authors are thankful to Dr. Kamira Weinfeld and Dr. Cecil Saguy, Surface Science Laboratory, Technion for the XPS and AFM measurements, respectively. The authors acknowledge the Department of Chemical Research Support, Weizmann Institute of Science, for single crystal XRD measurements. The authors are thankful to Dr. Yaron Kauffmann, Department of Materials Science & Engineering, Technion, for the STEM-HAADF measurements and Dr. Galit Atiya, Department of Materials Science & Engineering, Technion, for preparing samples for TEM using the FIB facility.  N. C. thanks Shahar Zuri from the Technion, for stimulating discussions.

**Conflict of Interest**

Authors have no conflicts of interest to declare.

# Supporting Information

# Change in Magnetic Order in NiPS$_3$ Single Crystals Induced by a Molecular Intercalation

Nirman Chakraborty,[1] Adi Harchol,[1] Azhar Abu-Hariri[1], Rajesh Kumar Yadav,[2,3] Muhamed Dawod,[4] Diksha Prabhu Gaonkar,[1] Kusha Sharma,[1] Anna Eyal,[5] Yaron Amouyal,[4] Doron Naveh[2,3] and Efrat Lifshitz[1*]

[1]*Schulich Faculty of Chemistry, Solid State Institute, Russel Berrie Nanotechnology Institute, Grand Program for Energy and the Helen Diller Quantum Center, Technion-Israel Institute of Technology, Haifa 3200003, Israel*

[2]*Faculty of Engineering, Bar-Ilan University, Ramat-Gan 5290002, Israel*

[3]*Institute of Nanotechnology and Advanced Materials, Bar-Ilan University, Ramat-Gan 5290002, Israel*

[4]*Department of Materials Science and Engineering, Technion-Israel Institute of Technology, Haifa 3200003, Israel*

[5]*Department of Physics, Technion, Haifa 3200003, Israel*

*Corresponding author: ssefrat@technion.ac.il

**Contents**

# 1. Crystal composition and stoichiometry

**Table S1.** Quantitative data of elemental analysis of different spots (spot 1 and spot 2) on pristine $NiPS_3$ sample (see Figure S1 and S2) using SEM-EDX. The formula units and sulphur vacancy/formula unit are provided in orange colour rows.

| Spot 1 | | | | | | |
|---|---|---|---|---|---|---|
| Element | Line Type | Apparent Concentration | k Ratio | Wt% | Wt% Sigma | Atomic % |
| P | K series | 70.00 | 0.39152 | 16.73 | 0.16 | 20.18 |
| S | K series | 133.76 | 1.15233 | 50.76 | 0.29 | 59.14 |
| Ni | K series | 80.80 | 0.80799 | 32.51 | 0.35 | 20.68 |
| Total: | | | | 100.00 | | 100.00 |
| | Formula | $NiPS_{2.88}$ | Sulphur vacancy | 4% | | |
| Spot 2 | | | | | | |
| P | K series | 22.22 | 0.12427 | 16.12 | 0.20 | 19.84 |
| S | K series | 41.83 | 0.36036 | 47.62 | 0.36 | 56.61 |
| Ni | K series | 30.39 | 0.30386 | 36.26 | 0.44 | 23.55 |
| Total: | | | | 100.00 | | 100.00 |
| | Formula | $NiPS_{2.85}$ | Sulphur vacancy | 5% | | |

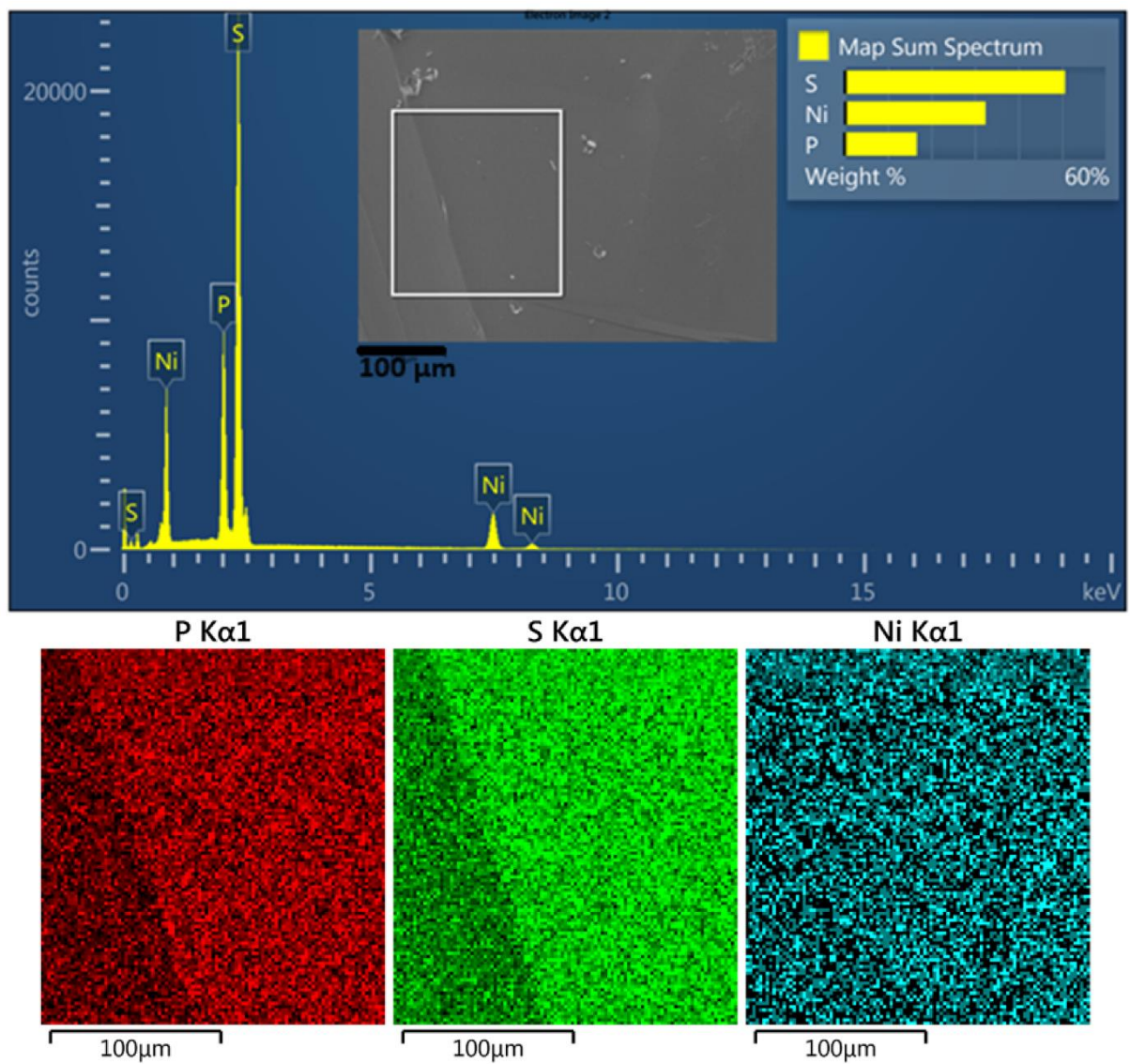

**Figure S1.** SEM-EDX spectrum of spot 1 on pristine NiPS$_3$ sample. The inset shows the SEM image of the spot of analysis. The individual elements are shown as color maps.

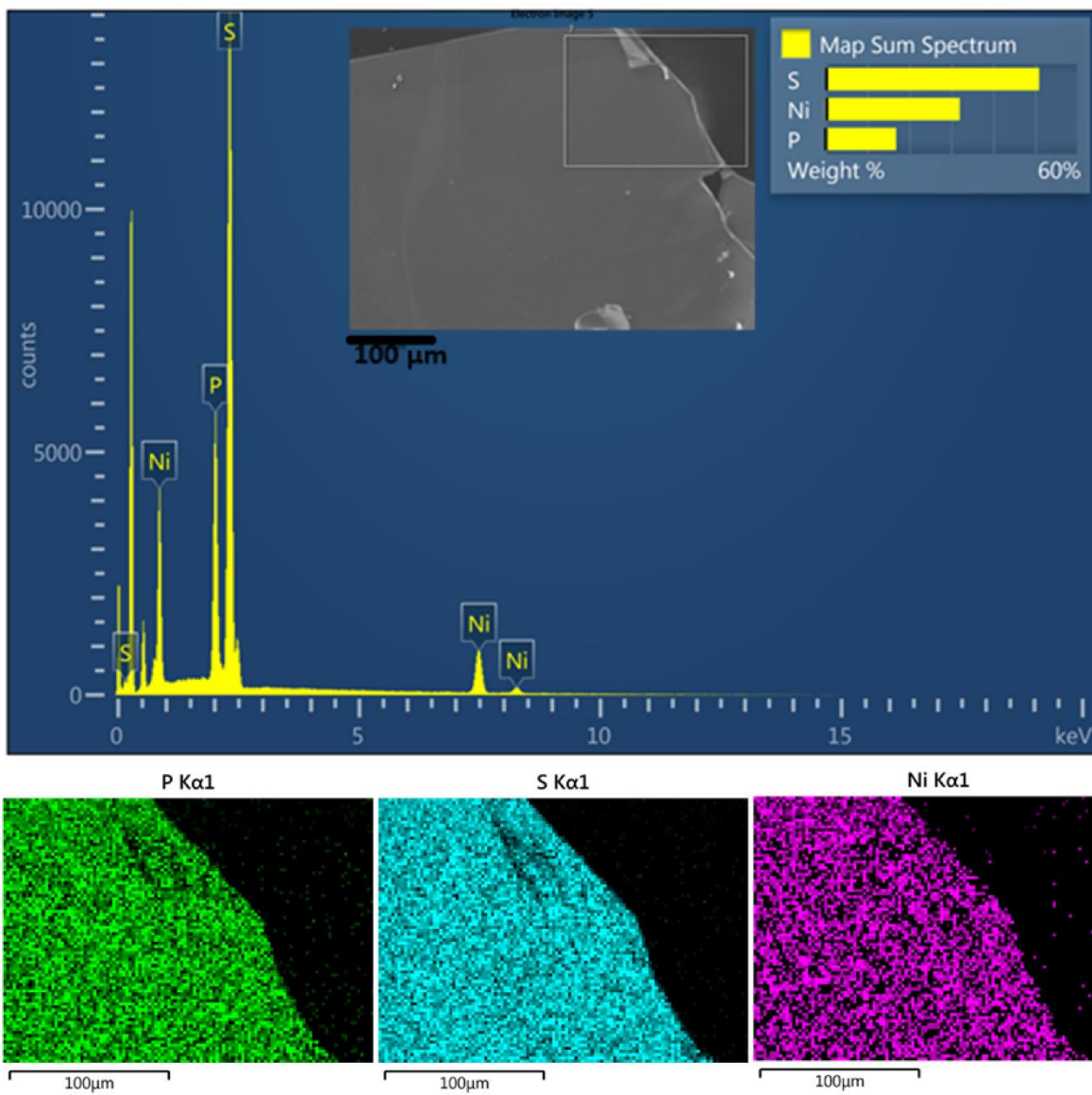

**Figure S2.** SEM-EDX spectrum of spot 2 on pristine NiPS$_3$ sample with individual elements shown as color maps. The inset shows the SEM image of the spot of analysis.

## 2. SXRD measurements and data analyses

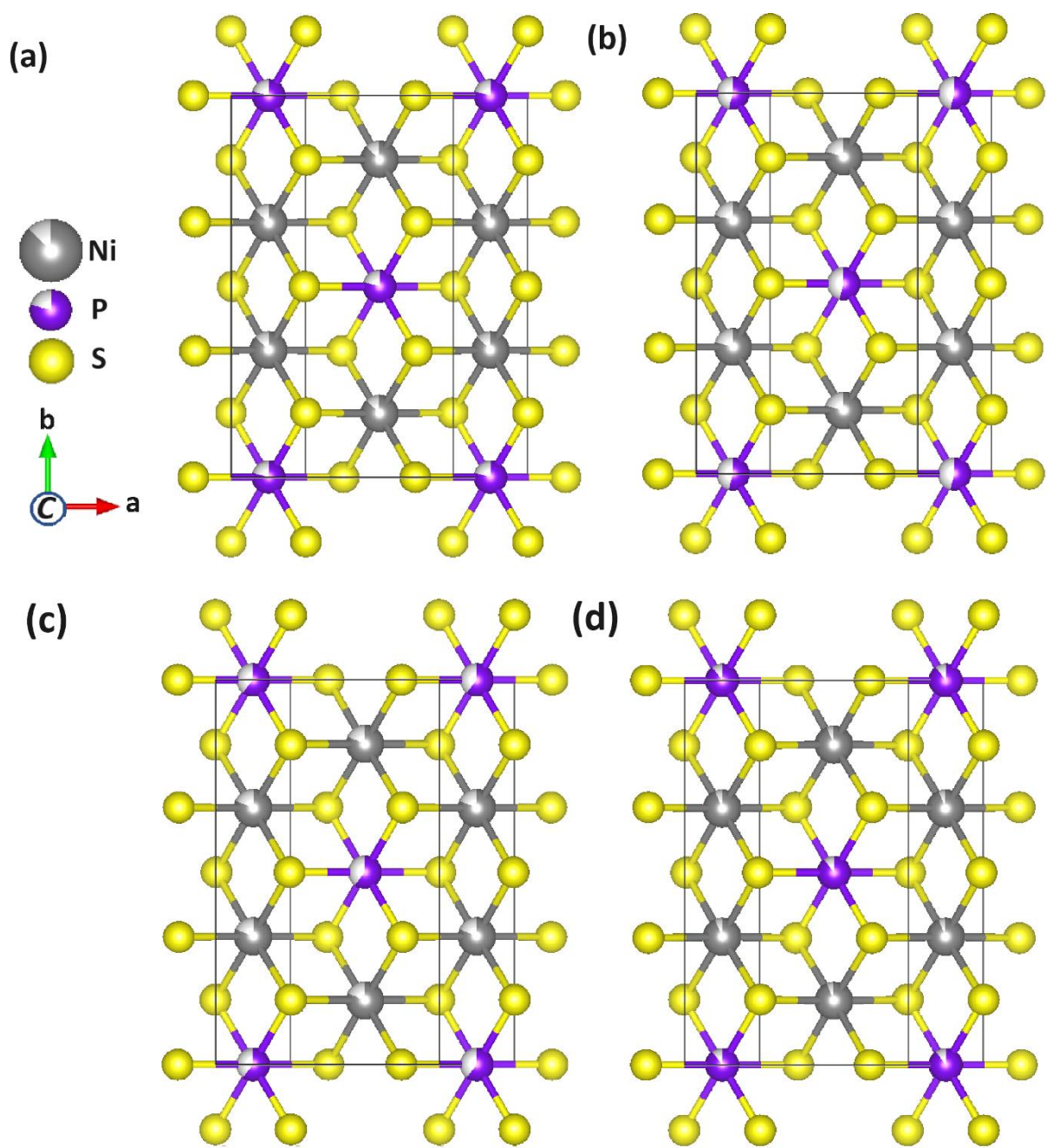

**Figure S3.** (a-d) A top view of the crystal structure obtained from SXRD measurement at 100 K, representing a unit cell (black frame) of PY_25, PY_40, PY_50, and PY_65, respectively. The plots refer to Table S2 below with the difference in bond lengths, as obtained from the highlighted crystal structures. The light gray shades in dark grey and purple spheres represent the average disorder in atomic site occupancies (see results and discussion in the main manuscript).

**Table S2.** Bond lengths (Å) obtained from SXRD of the pristine and intercalated $NiPS_3$ crystals.

| Bond | $NiPS_3$ | $NiPS_3\_PY\_25$ | $NiPS_3\_PY\_40$ | $NiPS_3\_PY\_50$ | $NiPS_3\_PY\_65$ |
|---|---|---|---|---|---|
| Ni1-S1(001) | 2.4553(9) | 2.4631(14) | 2.4684(10) | 2.4613(10) | 2.4583(12) |
| Ni1-S2(-100) | 2.4593(8) | 2.4624(10) | 2.4635(8) | 2.4633(8) | 2.4615(10) |
| Ni1-S2(111) | 2.4593(8) | 2.4624(10) | 2.4635(8) | 2.4633(8) | 2.4615(10) |
| Ni1-S1(000) | 2.4553(9) | 2.4631(14) | 2.4684(10) | 2.4613(10) | 2.4583(12) |
| Ni1-S1(111) | 2.4579(7) | 2.4642(10) | 2.4637(9) | 2.4613(9) | 2.4585(10) |
| Ni1-S1(010) | 2.4579(7) | 2.4642(10) | 2.4637(9) | 2.4613(9) | 2.4585(10) |
| P1-S1(111) | 2.0149(9) | 2.0115(15) | 2.0101(11) | 2.0168(10) | 2.0255(13) |
| P1-S1(101) | 2.0149(9) | 2.0115(15) | 2.0101(11) | 2.0168(10) | 2.0255(13) |
| P1-S2(111) | 2.1056(17) | 2.008(3) | 2.008(3) | 2.0224(18) | 2.030(3) |
| P1-P1(000) | 2.1640(3) | 2.174(4) | 1.082(2) | 2.153(3) | 2.162(4) |

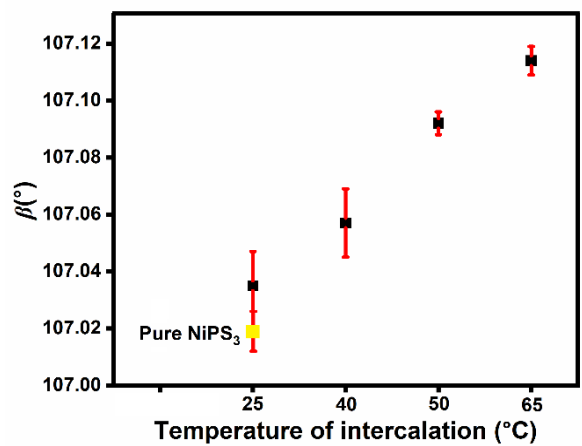

**Figure S4.** Variation of "$\beta$" with temperature of intercalation. The standard uncertainties are in red bars. Pristine $NiPS_3$ has been represented by the yellow square.

## 3. Surface morphology and cross-sectional measurements

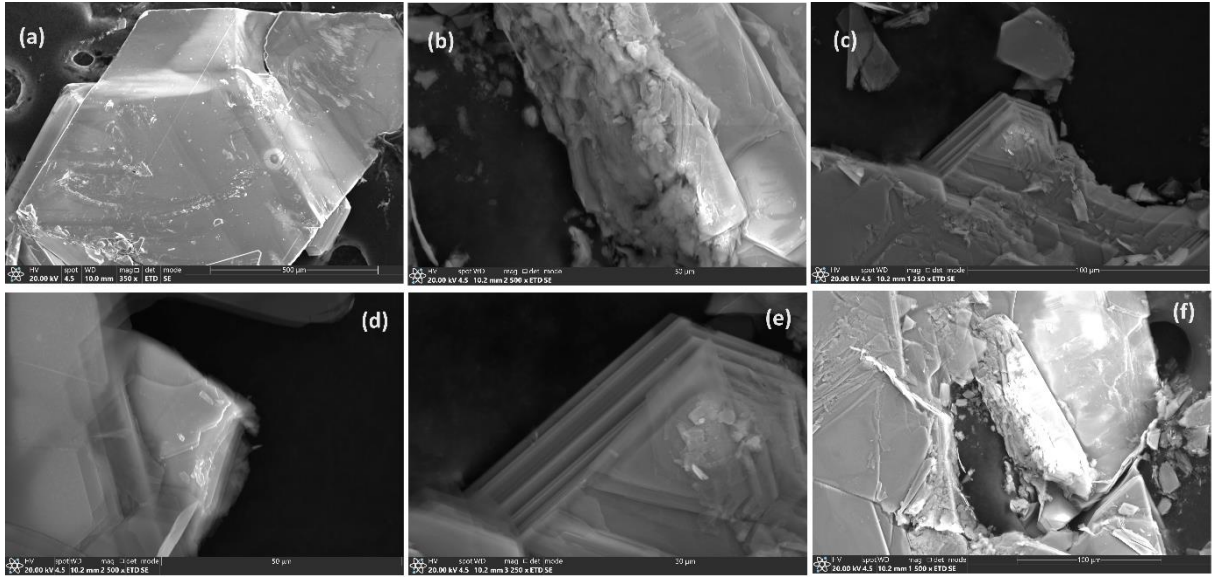

**Figure S5.** SEM images of (a) Pristine NiPS$_3$; (b-f) PY_65, when using backscattered electrons. The damage to the sample after intercalation, in addition to the loss of crystallinity, is evident from the above images.

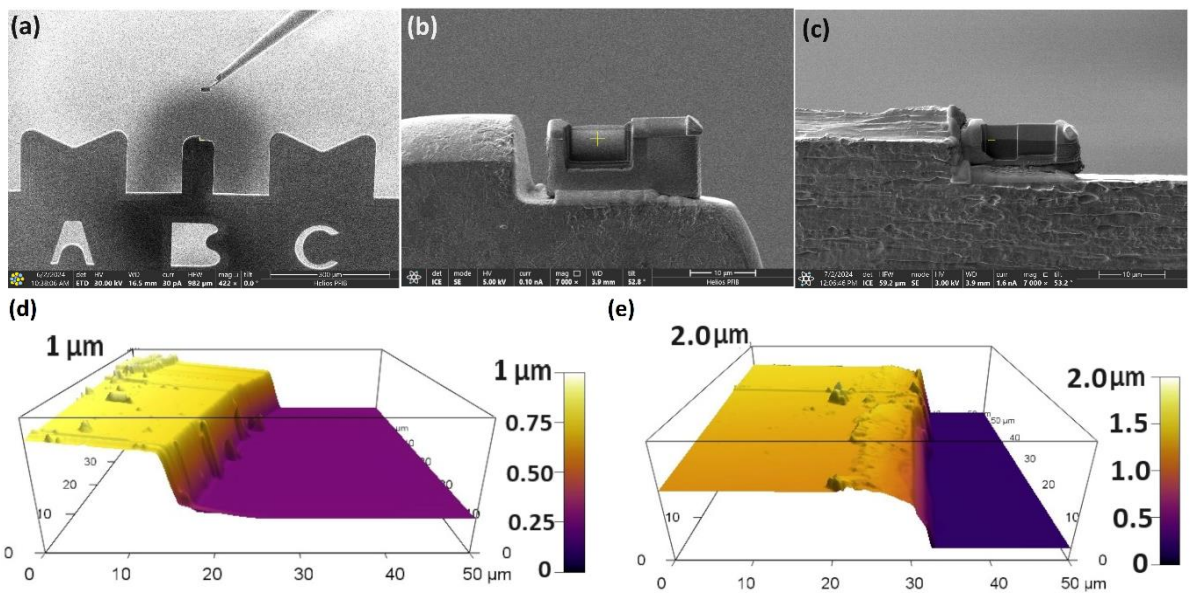

**Figure S6.** Images of extraction of cross-section of the single crystals using focused ion beam (FIB) technique. (a) The tip lifts off the material after being cut by an ion beam. (b, c) Freshly

cut cross-sections of NiPS$_3$ and PY_65, respectively. (d, e) Thickness profile of NiPS$_3$ and PY_65, respectively, using AFM in tapping mode. The measurement has been done with the surface of Si wafer (on which the crystal was placed) as a baseline. The average thickness of pristine NiPS$_3$ was estimated to ~1 μ and that for PY_65 was ~1.6 μ.

## 4. XPS and Raman data

**Table S3.** Absolute binding energy (B.E.) (in eV) of core-level spectra of different elements in pristine and intercalated NiPS$_3$ samples from XPS measurements.

| Sample | NiPS$_3$ | PY_25 | PY_40 | PY_50 | PY_65 |
|---|---|---|---|---|---|
| Ni 2P$_{3/2}$ Ni$^{2+}$/Ni$^{3+}$ | 853.94/855.84 | 853.98/856.28 | 854.10/856.40 | 854.14/856.44 | 854.25/856.55 |
| S 2P$_{3/2}$/S 2P$_{1/2}$ | 161.76/162.96 | 161.91/163.11 | 162.08/163.28 | 162.11/163.31 | 162.26/163.46 |
| P 2P$_{3/2}$/S 2P$_{1/2}$ | 131.16/132.06 | 131.17/132.07 | 131.28/132.18 | 131.49/132.39 | 131.66/132.56 |
| C-N/N-H$^+$ | ─── | 400.11/402.01 | 400.00/401.90 | 399.98/401.88 | 399.96/401.86 |

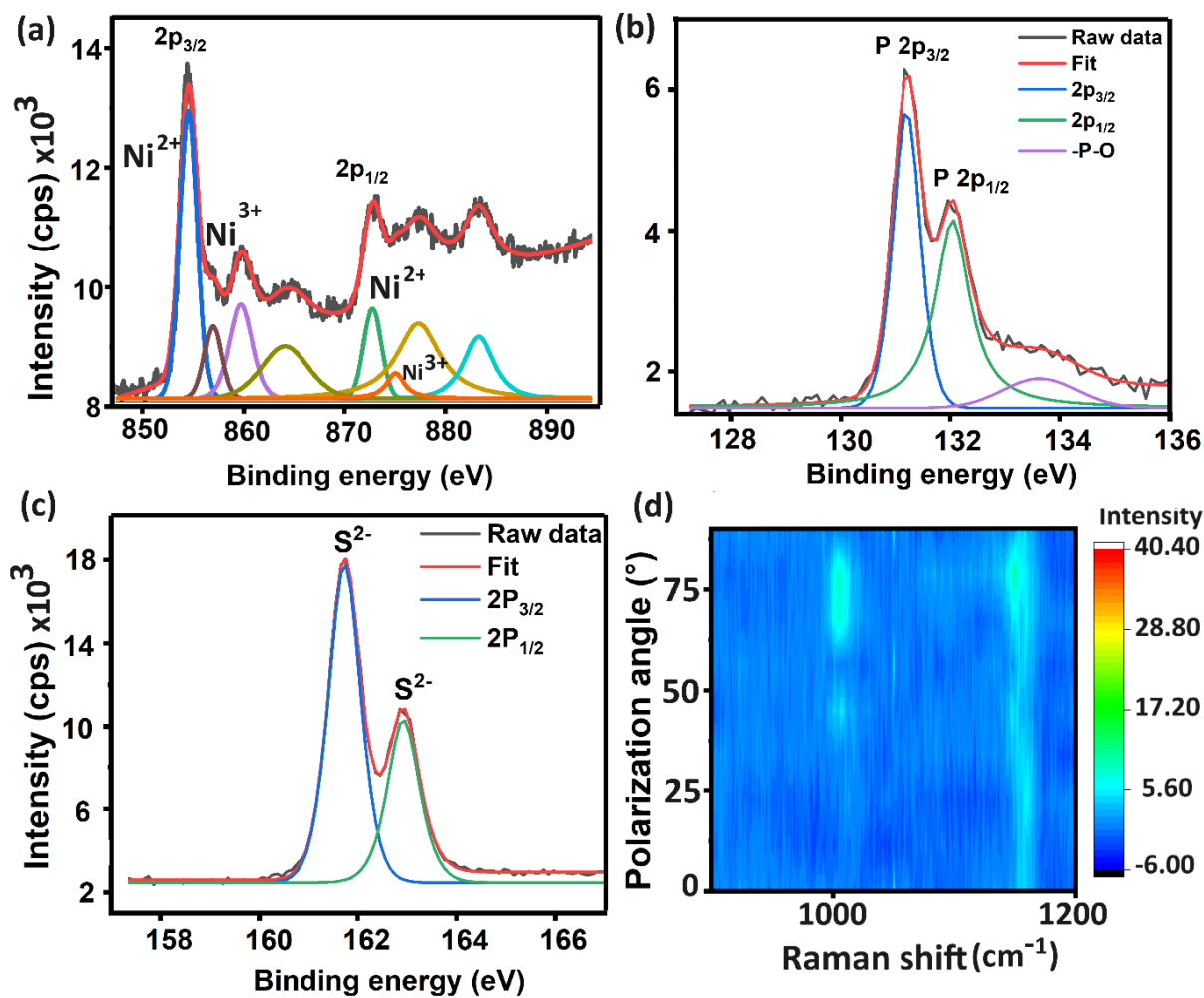

**Figure S7.** (a-c) Ni 2p, P 2p, S 2p XPS core-level spectra of PY_65. (d) Raman spectrum of pyridine vibration at 1000 cm$^{-1}$ for PY_65 sample. Note that the signals at θ'=45°-50° and 65°-80° correspond to two different ranges of angles for orientation of the pyridine dipole (concerning the "*ab*" plane), θ=45°-50° and 50°-60°.

## 5. Polarization calculations from Raman data

Raman studies of all samples were conducted with a 532 nm laser at room temperature, incident normally on the "*ab*" plane of the sample. The procedure to measure the samples and conditions are elaborated in Methods. Figures S8a and S8b demonstrate the room temperature spectra of all samples. The inset of Figure S8b shows the scheme for Raman measurements and the used notation. The peaks show a slight red shift with increasing Δβ. As per reports, $E_g^1$ and $E_g^2$ modes have been reported to correspond to the vibrations, including heavy Ni atoms, while $A_{1g}^1$ and the $A_{1g}^2$ correspond to the intramolecular vibrations from $(P_2S_6)^{4-}$ bipyramidal structures [51-53]. The red shift indicates the expansion of the Ni-S and P-S bonds, which is in conformation with the trends in bond lengths obtained from SXRD measurements (see Table

S2). Figures S8c and S8e demonstrate the polarization-dependent Raman spectra at room temperature, highlighting the $E_g^1$ and $E_g^2$ modes for NiPS$_3$ and representative sample PY_40, respectively. Figures 5d and 5f highlight the $E_g^3$, $A_{1g}^1$ and $A_{1g}^2$ modes. Investigating the change in Raman modes revealed a significant variation in the intensity with polarization angle. While the degree of linear polarization of the $E_g^1$ and $E_g^2$ modes in NiPS$_3$ were 12% and 17%, respectively, the same increased to 23% and 31% in NiPS$_3$_PY_40. The $A_{1g}^2$ mode exhibited a doubling in the degree of linear polarization, indicating the strong influence of the pyridine dipole on the local bonds on either side of the vdW gap. This indicates the generation of polarity within the intercalated crystals, which were supplemented by room-temperature capacitance measurements. The details of capacitance measurements and calculations are provided in section 6.

The degree of linear polarization (DLP) for different Raman modes was calculated using the formula

$DLP = \frac{I_{max} - I_{min}}{I_{max} + I_{min}}$ …………(1), where $I_{max}$ and $I_{min}$ are the maximum and minimum intensities, respectively. The intensity values were extracted after fitting the individual Raman curves using a single Gaussian function.

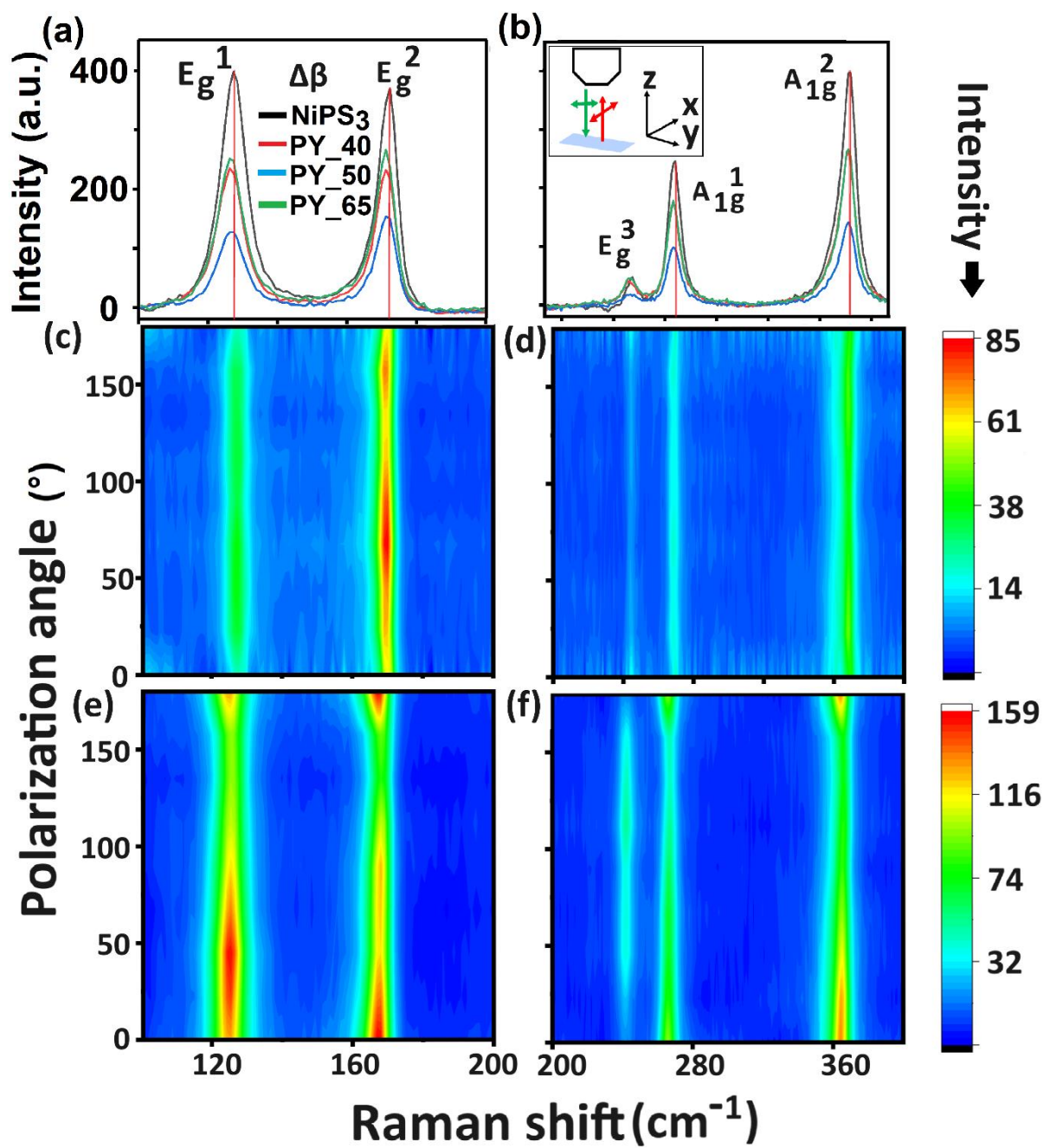

**Figure S8. Raman measurements.** (a, b) Room temperature Raman spectra collected for pure and intercalated NiPS$_3$ samples. The red bar is used as a reference to indicate a shift in the peak of Raman intensity with $\Delta\beta$. Inset of (b) shows the schematic of the geometry of measurement used, involving the microscope head, incident (green) and emitted(red) light and corresponds to the $z(yx)\bar{z}$ notation. Polarization-dependent room temperature Raman spectra of (c, e) pristine NiPS$_3$ and (d, f) PY_40, respectively.

## 6. Capacitance and conductivity measurements

To fabricate the pristine and intercalated-NiPS$_3$ device, a polydimethylsiloxane (PDMS)-based dry transfer method was performed within a glovebox to prevent contamination and degradation (see Figure S9a and S9b). Si/SiO$_2$ substrates were cleaned with deionized water, acetone, and isopropanol and then dried under nitrogen flow. Back electrodes (Ti/Au: 5/30 nm) were patterned on these substrates using E-beam writing and metal deposition (see Figure S9a and S9b). Mechanical exfoliation was used to obtain intercalated-NiPS$_3$ flakes of the desired thickness from the bulk crystals onto PDMS. Using the PDMS-based dry transfer technique, the flakes were transferred onto the freshly fabricated back electrode. To make the device behave as a parallel plate capacitor, the top electrode (Ti/Au: 5/30 nm) was patterned over the transferred flakes using E-beam writing and metal deposition (see Figure S9a and S9b). The thickness of the flakes was measured using AFM (see Figure S9d and S9f). Capacitance-Voltage (CV) and conductivity measurements were performed simultaneously using a Keysight B1500A Semiconductor Device Analyzer. The semiconductor device sample was mounted inside a Linkam stage to ensure precise Room temperature (25 °C) control during the measurements. The Linkam stage was then connected to the B1500A, and electrical contacts were established using a wire-bonder. The Keysight EasyEXPERT software was used to configure the measurement parameters, with a voltage sweep ranging from -5 V to 5 V and a measurement frequency of 1 MHz. Following the setup, the CV measurements were conducted by applying the voltage sweep and recording the capacitance at each step (see Figure S9g and S9h).

The permittivity of pristine and intercalated NiPS$_3$ samples was calculated from the concept of capacitance measurements of a parallel-plate capacitor configuration using the formula $C = \varepsilon \frac{A}{d}$ ………..(2), where C is the capacitance measured, $\varepsilon$ is the permittivity of the medium between two plates, "d" is the distance between two plates of the capacitor and "A" is the area. For the present case, the C value has been extracted from the C vs V plots (see Figure S9g and S9h). "d" has been calculated to be 63 nm and 43 nm for pristine and intercalated NiPS$_3$ samples using AFM in tapping mode, respectively (see Figure S9d and S9f). The area of the parallel plate region has been measured to be 3.9×10$^{-11}$ m$^2$ and 6.0×10$^{-11}$ m$^2$ for pristine and intercalated NiPS$_3$ samples, respectively. Using the above values, $\varepsilon$ for pristine NiPS$_3$ was calculated to be 2.45×10$^{-8}$ F/m and 3.72×10$^{-8}$ F/m for intercalated NiPS$_3$ (PY_40). The conductivity of pristine $NiPS_3$ was $1.6\times10^{-6}$ $\Omega^{-1}$, that for PY_40 and PY_50 was $1.5\times10^{-5}$ $\Omega^{-1}$ and $3.2\times10^{-5}$ $\Omega^{-1}$, respectively.

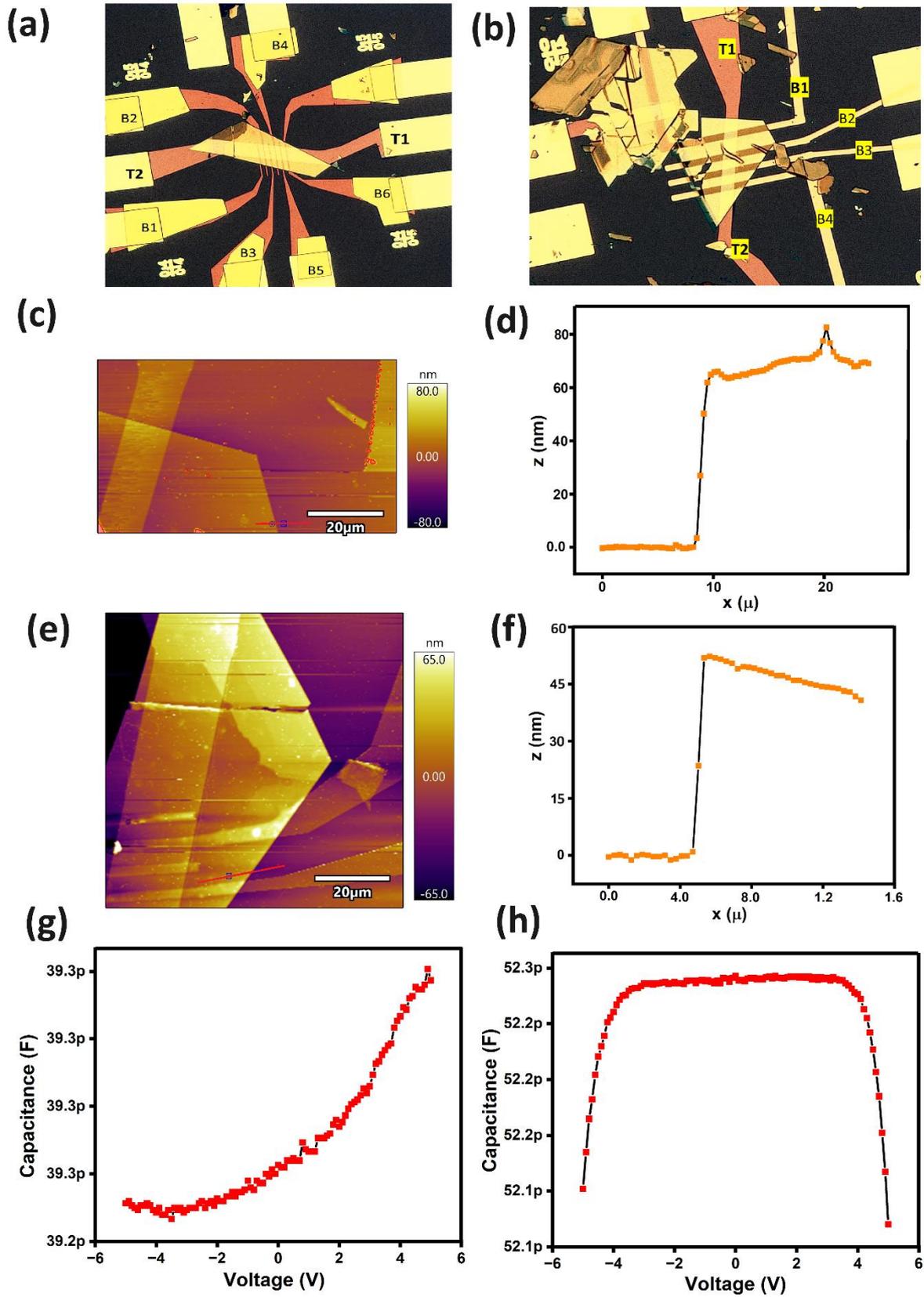

**Figure S9.** (a) Optical microscopic images of devices made in parallel-plate configuration for pristine NiPS$_3$ (b) PY_40. (c, d) AFM topography and profile thickness of pristine NiPS$_3$ (e, f) PY_40, collected in tapping mode. (g, h) C vs. V measurements of NiPS$_3$ pristine and PY_40, respectively, with source frequency of 1 MHz.

## 7. Magnetic susceptibility measurements and determination of magnetic phase transition from Raman spectroscopy

Magnetic susceptibility was calculated from magnetization data, using the formula:

$$M = \chi_M H \quad \ldots\ldots\ldots\ldots(3)$$

where M=magnetization, $\chi_M$=magnetic susceptibility and H=external field.

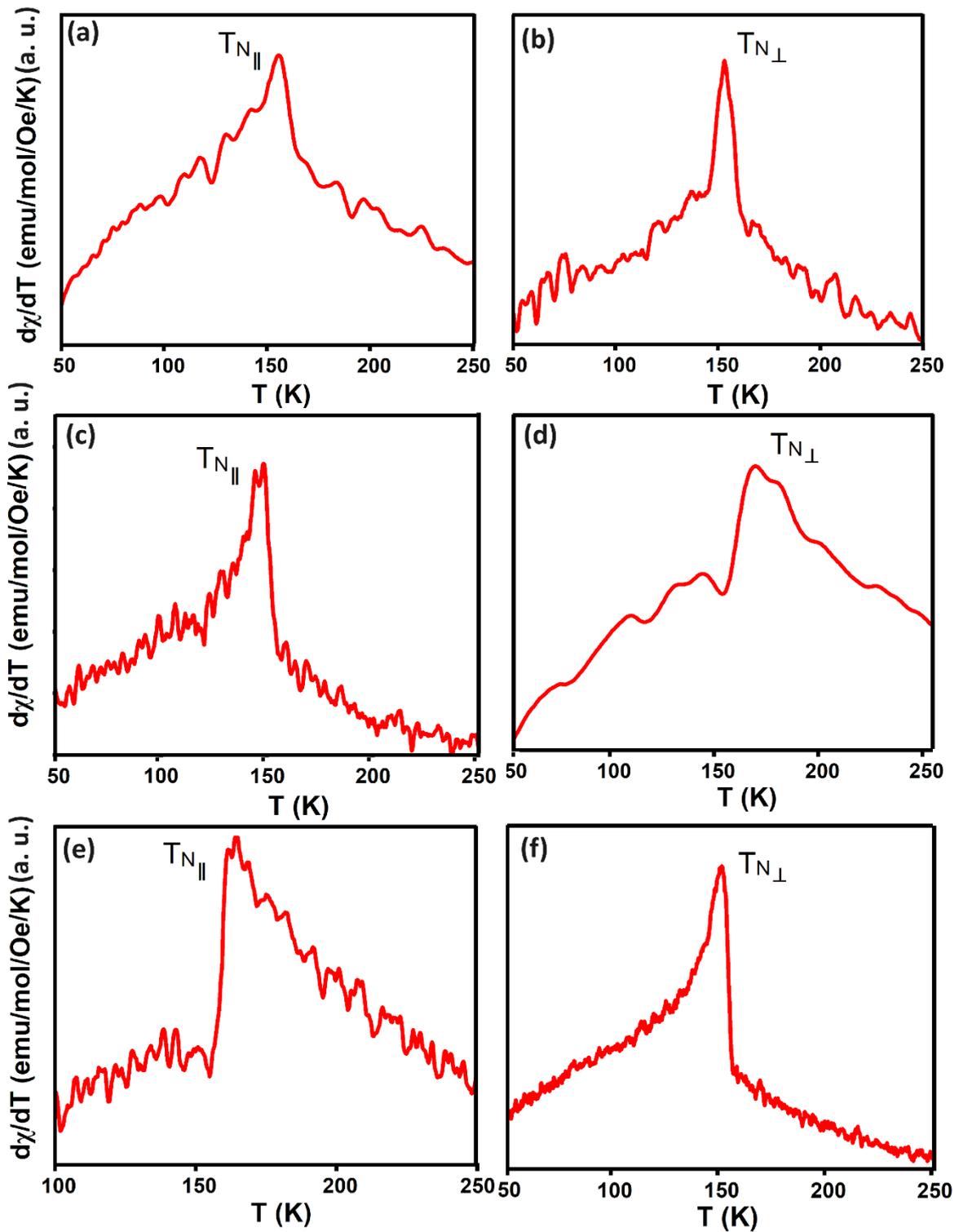

**Figure S10.** dχ/dT vs. T curves for (a, b) pristine $NiPS_3$ (c, d) PY_40 (e, f) PY_65. The Néel temperature has been estimated from the singularity observed.

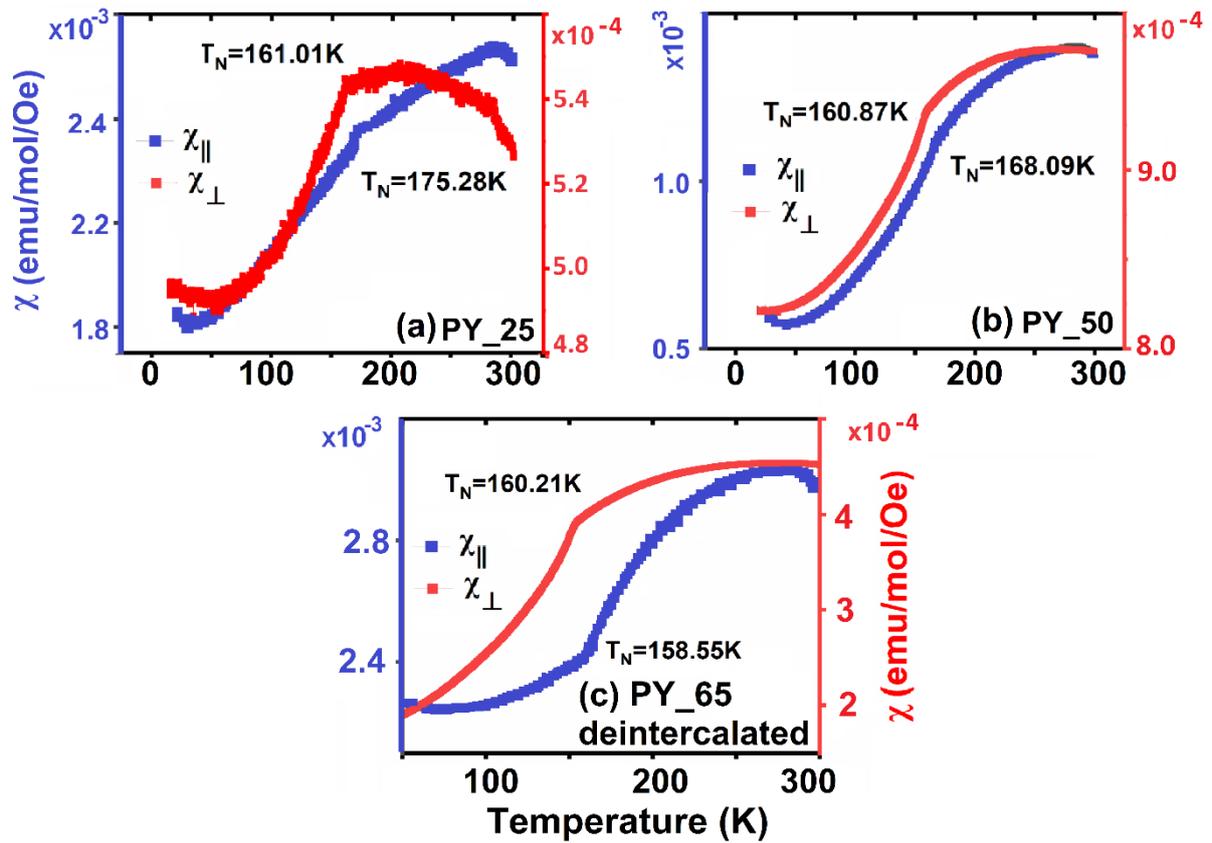

**Figure S11.** Magnetic susceptibility (χ) vs. temperature curve for (a) PY_25, (b) PY_50 and (c) PY_65 after deintercalation.

Trends in magnetic phase transitions obtained from SQUID measurements were further validated using temperature-dependent micro-Raman studies [35]. It has been reported that the $E_g^2$ mode for NiPS$_3$ consists of two nearly degenerate components, observed separately in cross-polarization configurations [5]. The onset of spin-phonon coupling is identified by the AFM spin-ordering temperature [5]. Thus, the variation of the $E_g^2$ mode with temperature, under the $z(y\bar{x})\bar{z}$ and $z(yx)\bar{z}$ configurations of polarization (see inset of Figure S8(b)), can delineate the onset of AFM ordering due to spins oriented in one direction on the "*ab*" plane and the additional spin components (including out-of-plane), respectively [57]. The spectra were collected under excitation with a 532 nm laser, incident normally on the "*ab*" plane, with sample temperature varying from 130 K to 200 K.

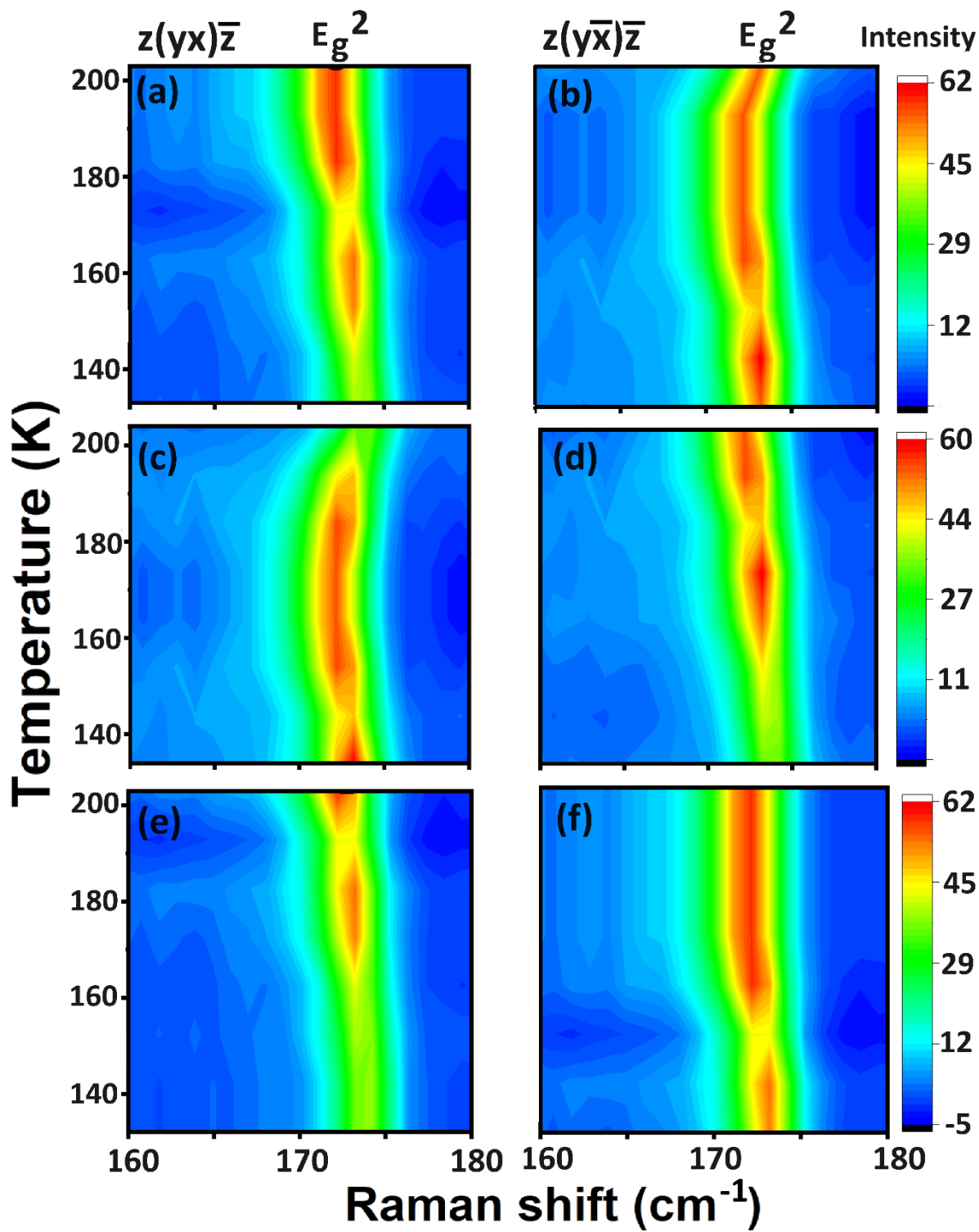

**Figure S12. Phase change analyses from temperature-dependent Raman studies.** Temperature-dependent variation of the $E_g^2$ Raman mode for pristine and intercalated NiPS$_3$ samples PY_40 and PY_65, respectively (a-e) in $z(yx)\bar{z}$ configuration (b-f) in $z(y\bar{x})\bar{z}$ configuration. Note that the phase transition is visualized in the form of discontinuity in the spectral maxima.

## 8. Thermal property measurements

**Table S4.** Fitting parameters (in arbitrary units) of 3-dimensional heat flow using a 3-variable ellipsoidal fitting equation.

| $\Delta\beta$ | A | $A_0$ | B | $B_0$ | C | $C_0$ |
|---|---|---|---|---|---|---|
| 0.016 | 0.3172(4) | 0.1561 | 0.5411(3) | 0.2133 | 0.2565(4) | 0.1218 |
| 0.038 | 0.4981(5) | 0.3444 | 0.3104(5) | 0.1432 | 0.3230(3) | 0.2111 |
| 0.095 | 0.1027(1) | 0.4322 | 0.2921(7) | 0.1123 | 0.7332(8) | 0.8871 |